\documentclass[aps,preprint,showpacs,floats,epsf,epsfig,nofootinbib,12pt]{revtex4}
\usepackage{graphicx}
\usepackage{epsfig}
\usepackage[dvips]{color}

\def\beq{\begin{eqnarray}}
\def\eeq{\end{eqnarray}}
\def\non{\nonumber}

\def\la{\langle}
\def\ra{\rangle}
\def\Mb{M_{\Sigma_b}}
\def\Mc{M_{\Sigma^*_c}}

\begin{document}

\title{  $\Sigma_{b}\to\Sigma_c^*$ weak decays in the light-front quark model with two schemes to deal with the polarization of diquark}

\vspace{1cm}

\author{ Hong-Wei Ke$^{1}$   \footnote{khw020056@hotmail.com}, Ning Hao$^{1}$ and
        Xue-Qian Li$^2$\footnote{lixq@nankai.edu.cn},
   }

\affiliation{  $^{1}$ School of Science, Tianjin University, Tianjin 300072, China
\\
  $^{2}$ School of Physics, Nankai University, Tianjin 300071, China }

\vspace{12cm}

\begin{abstract}
Thanks to the remarkable achievements of LHC,
a large database on baryons
has been accumulated, so it is believed that the time for
precisely studying baryons
especially heavy baryons, has
come. By analyzing the data, the quark-diquark structure which has
been under intensive discussions, can be tested. In this work the
decay widths of weak transitions $\Sigma_b\to \Sigma^*_c+X$ are
calculated in terms of the light front quark model (LFQM). To
carry out the calculations, the
quark-diquark  picture is employed where an axial-vector diquark
composed of two light quarks serves as a spectator in the
concerned processes.  The first step of this work is to construct
the vertex functions for $\Sigma^{(*)}_c$ and $\Sigma_b$, then the
relevant form factors are derived. It is shown that
under the heavy quark limit the
Isgur-Wise functions for the transition are re-deduced. Indeed, how
to properly depict the polarization ($\epsilon_{\mu}$) of the
diquark is slightly tricky. In this work, we apply two schemes to
explicitly determine the momentum-dependence of the diquark. The
corresponding numerical results are presented which will be
testified by the future experiments.


\end{abstract}

\maketitle

\section{Introduction}
It is well noted, the data of mesons  much exceed those of
baryons, so the studies on the mesonic properties such as the mass
spectra, production and decay mechanisms are abundant as well as
many so called anomalies in the wide area. The reason is obvious
that meson productions are more easily realized for
electron-positron colliders. By contrast, the statistics on
baryonic properties is relatively poor. Baryons possess three
valence constituents  whereas there are only two (quark-antiquark)
in regular mesons. From the theoretical aspect, dealing with a
three-body system is much more complicated than a two-body
one. This is an old problem even exists in classical mechanics.
Indeed, when a sufficiently large
database lacked, an accurate
study on baryons seemed not so compelling yet.

However at present, the situation is changed by emergence of
great number of baryons, especially those containing one or two
heavy quarks (anti-quarks). Researchers have opportunities to
carry out more accurate studies on the baryonic processes. The standard
approach for handling three-body systems is the Faddeev equations,
which mathematically is rather difficult to be solved unless some
simplification is taken into account.  Among all the
simplification schemes, the quark-diquark picture is more
realistic especially for the heavy baryons. In the weak decay
of a heavy baryon into another heavy baryon, only the heavy quarks take part in
the transition while
the light quarks can be regarded as spectators. At least at
the leading order this is a good approximation. Thus the spins,
flavors and isospins of the two light quarks remain as conserved
quantum numbers during the transition, so that their combination
can be considered as a loosely bound subsystem,
i.e. a diquark whose inner structure can be manifested by a form
factor. Generally, the form factor possesses a few free parameters
which would be determined by fitting data or via some
model-dependent calculations. The picture of baryon is most
commonly adopted for calculating the mass spectra and transition
rates. The authors of
Refs.\cite{Korner:1992uw,Ebert:2006rp,Singleton:1990ye,Ivanov:1996fj,Ivanov:1998ya,Wei:2009np}
studied the transitions between two heavy baryons in terms of the
quark-diquark picture and the theoretical results are
qualitatively accordant with the data. The picture is still in
acute dispute although it is very effective and useful for studying
the processes where baryons are involved. To confirm the validity
of the picture, one may study typical transitions between heavy
baryons. That is the goal of our paper.

In this paper we will study the transition $\Sigma_{b}\rightarrow
\Sigma^*_{c}$  where the light diquark in the decay  is regarded
as a spectator. The hadronic matrix element for
$\Sigma_{b}\rightarrow \Sigma^*_{c}$ is generally parameterized by
a few form factors. Under the heavy quark limit\cite{IW} those
form factors can be reduced into two Isgur-Wise
functions\cite{Korner:1992uw,Ebert:2006rp}. Some authors
\cite{Korner:1992uw,Ebert:2006rp,Singleton:1990ye,Ivanov:1996fj,Ivanov:1998ya}
calculated the form factors for $\Sigma_{b}\rightarrow
\Sigma^*_{c}$ in various approaches where heavy quark limit was
employed. Here we may consider a more general case beyond the
heavy quark effective theory (HQET) and obtain the form factors.
Parallel to numerous phenomenological models, the light-front
quark model (LFQM) has been successfully applied to study
transitions between different mesons
\cite{Jaus,Ji:1992yf,Cheng:1996if,Cheng:2003sm,Hwang:2006cua,Li:2010bb,Ke:2009ed,Wei:2009nc,Choi:2007se}.
{For example in Ref.\cite{Cheng:2003sm} the authors predicted some
decay rates of $B$ and the results are consistent with data (Table XIV in
Ref.\cite{Cheng:2003sm}). The decay constants of several charmed
mesons are calculated in Ref.\cite{Choi:2007se} and are accordant
to the data (Table II In Ref.\cite{Choi:2007se}).} The approach
has also been applied to deal with transition amplitudes between
two baryons in the quark-diquark
picture\cite{Ke:2007tg,Ke:2012wa}.

However the formulas in
\cite{Ke:2012wa} cannot be directly applied to
$\Sigma_{b}\rightarrow \Sigma_{c}^*$ because the quantum number of
$\Sigma_{c}^*$ is
$\frac{3}{2}^+$\cite{Korner:1992uw,Ebert:2006rp}. Since the
isospin of both $\Sigma_b$ and $\Sigma_c^{(*)}$ is 1, the light $ud$
diquark must be in an isospin-1 and color $\bar 3$ state. To
guarantee the wavefunction of the $ud$ diquark to be totally
antisymmetric for spin$\times$color$\times$flavor the spin of the
$ud$ system should be 1. Thus we need to re-construct the vertex
function for a $\frac{3}{2}^+$ heavy baryon which is regarded as a
bound state of a heavy quark and a light axial vector diquark. The
key problem is to validate the momentum dependence of the
polarization of the axial diaquark. Naively, one should expect
that the polarization vector in the vertex function directly
depends on the momentum of the diquark. While the vertex function
containing the polarization vector is deduced from the
Clebsch-Gordan (CG) coefficients, the polarization uniquely
depends on the total momentum of the baryon. In order to study the
vertex function with an axial diquark we employ two schemes to
deal with the polarization of the axial vector. We name them as
Scheme I where the polarization vector depends on the momentum of
the baryon and Scheme II where the polarization vector depends on
the momentum of the diquark. Then using the Feynman rules in LFQM we
write down the transition matrix element which is parametrized by
eight form factors. Under the heavy quark limit, the transition
matrix element of $\Sigma_b\to\Sigma^*_c$ is reduced into a simple
form which is described  by merely two generalized Isgur-Wise
functions.

In this paper we will explore the the semileptonic and non
leptonic decays of $\Sigma_b\to\Sigma^*_c$. By studying these
decays of baryons we can get a better understanding on baryon
structures and interactions among the constituents. These decays
also provide valuable information of the CKM parameters and serve
as an ideal laboratory to study non-perturbative QCD effects in
the heavy baryon system. The semileptonic decay is relatively
simple and not contaminated by the final state re-scattering
effects, therefore one only needs to consider the pure hadronic
transition between two hadrons, thus
studies on semileptonic decays might be more favorable for
testing the employed model and/or constrain the model parameters.
Indeed, comparing our theoretical results with data the model
parameters which are hidden in the vertex functions can be fixed.
Even the widths of the non-leptonic decay $\Sigma_{b}\rightarrow
\Sigma^*_{c}+M$ can also be evaluated in a similar way while
assuming the interaction between the produced meson and heavy baryon
to be weak, so that final state interactions can be
ignored. Within the framework of the light-front model some
parallel approached developed in
Ref.\cite{Ma:2002ir,Ma:2002xu,Brodsky:1985gs,Brodsky:2016yod} were
employed to study the decay of hadron.

This paper is organized as follows: after the introduction, in
section II we construct the vertex functions of heavy baryons, then
write down the transition amplitude for $\Sigma_{b}\rightarrow
\Sigma^*_{c}$ in the light-front quark model and deduce the form
factors with (without) using the heavy quark approximation, then we
present our numerical results for $\Sigma_{b}\rightarrow
\Sigma^*_{c}$ along with all necessary input parameters in section
III. Section IV is devoted to our conclusion and discussions.

\section{$\Sigma_{b}\rightarrow \Sigma_{c}^*$ in the light-front quark model}

Assuming the quark-diquark structure\cite{Korner:1992uw,Ebert:2006rp},
the heavy baryons $\Sigma_{b}$, $\Sigma^*_{c}$ and $\Sigma_{c}$
consist of a light $1^+$ diquark [ud] and one heavy quark $b(c)$.
To ensure the right  quantum numbers of $({1\over 2})^+$ and
$({3\over 2})^+$, the orbital angular momentum between the two
components is zero, i.e. $l=0$, while the spin of the diquark is
1. First we need to construct the vertex functions of $\Sigma_{Q}$
and $\Sigma^*_{Q}$ according to their quantum numbers. We employ
two schemes to deal with the polarization of the axial vector
for its momentum dependence.
\subsection{the vertex functions of $\Sigma_{Q}$ and
$\Sigma^*_{Q}$ in Scheme I}

In Refs.\cite{pentaquark1,pentaquark2} with the same model the
authors calculated decay rates of pentaquark which is supposed to be in
the antiquark-diquark-diquark structure. In their work appropriate
vertex functions are constructed. Returning to our case where baryons
are of a quark-diquark structure, thus, a similarity between
baryon and pentaquark structures hints us that we may imitate the
way for constructing the vertex functions of
pentaquark\cite{pentaquark1,pentaquark2} to obtain the
corresponding those for
$\Sigma_Q$ and $\Sigma^*_Q$ . The wavefunction of $\Sigma_Q$ with
a total spin $S=1/2$ and momentum $P$ is
\begin{eqnarray}\label{eq:lfbaryon}
 |\Sigma_Q(P,S,S_z)\rangle&=&\int\{d^3\tilde p_1\}\{d^3\tilde p_2\} \,
  2(2\pi)^3\delta^3(\tilde{P}-\tilde{p_1}-\tilde{p_2}) \nonumber\\
 &&\times\sum_{\lambda_1,m}\Psi^{SS_z}(\tilde{p}_1,\tilde{p}_2,\lambda_1,\sigma)
  C^{\alpha}_{\beta\gamma}F^{ij}\left|\right.
  Q_{\alpha}(p_1,\lambda_1)[q_{1i}^{\beta}q_{2j}^{\gamma}(\sigma)](p_2)\ra,
\end{eqnarray}
where $Q$ represents $b$ or $c$, $[q_1q_2]$ here is $[ud]$,
$\lambda$ denotes its helicity, $\sigma$ stands for the polarization
projection, $p_1,~ p_2$ are the on-mass-shell light-front momenta
defined by
\begin{equation}
 \tilde{p}=(p^+,p_{\perp}),\qquad p_\perp=(p^x,p^y),\qquad
 p^-=\frac{m^2+p_{\perp}^2}{p^+},
\end{equation}
and
\begin{eqnarray}
&&\{d^3\tilde{p}\}\equiv\frac{dp^+d^2 p_{\perp}}{2(2\pi)^3},\qquad
  \delta^3(\tilde{p})=\delta(p^+)\delta^2(p_{\perp}),
  \nonumber\\
&&\mid Q(p_1,\lambda_1)[q_1 q_2](p_2)\rangle=
 b^{\dagger}_{\lambda_1}(p_1)a^{\dagger}(p_2)| 0\ra,\non\\
&&[a_{\sigma'}(p'),
a_{\sigma}^{\dagger}(p)]=2(2\pi)^3\delta^3(\tilde{p}'-\tilde{p})\delta_{\sigma'\sigma},
  \nonumber\\
&&\{d_{\lambda'}(p'),d_{\lambda}^{\dagger}(p)\}=
  2(2\pi)^3\delta^3(\tilde{p}'-\tilde{p})\delta_{\lambda'\lambda},
\end{eqnarray}
The coefficient $C^{\alpha}_{\beta\gamma}$ is a normalized color
factor and $F^{ij}$ is a normalized flavor coefficient,
 \beq
 && C^{\alpha}_{\beta\gamma}F^{ij}C^{\alpha'}_{\beta'\gamma'}F^{i'j'}
  \la Q_{\alpha'}(p'_1,\lambda'_1)[q_{1i'}^{\beta'}q_{2j'}^{\gamma'}(\sigma')](p'_2)|
  Q_{\alpha}(p_1,\lambda_1)[q_{1i}^{\beta}q_{2j}^{\gamma}(\sigma)](p_2)\ra
  \non\\
  &&=2^2(2\pi)^6\delta^3(\tilde{p}_1'-\tilde{p}_1)\delta^3
  (\tilde{p}_2'-\tilde{p}_2)\delta_{\lambda'_1\lambda_1}\delta_{\sigma\sigma'}.
 \eeq

Intrinsic variables $(x_i, k_{i\perp})$ with $i=1,2$ are
introduced to describe the internal motion of the constituents
through
\begin{eqnarray}
&&p^+_1=x_1 \bar P^+, \qquad\qquad p^+_2=x_2 \bar P^+,
 \qquad\qquad x_1+x_2=1, \nonumber\\
&&p_{1\perp}=x_1 \bar P_{\perp}+k_{1\perp},
  ~~~ p_{2\perp}=x_2 \bar P_{\perp}+k_{2\perp},
  ~~~ k_{\perp}=-k_{1\perp}=k_{2\perp},
\end{eqnarray}
where $x_i$ are the light-front momentum fractions satisfing
$0<x_i<1$. The invariant mass square $M_0^2$ is defined as
 \begin{eqnarray} \label{eq:Mpz}
  M_0^2=\frac{k_{1\perp}^2+m_1^2}{x_1}+
        \frac{k_{2\perp}^2+m_2^2}{x_2}.
 \end{eqnarray}
The invariant mass $M_0$ is different from the hadron mass $M$
which satisfies the physical mass-shell condition $M^2=P^2$. This
is due to the fact that the baryon, heavy quark and diquark cannot
be on their mass shells simultaneously. The internal momenta
are defined as
 \beq
 k_i=(k_i^-,k_i^+,k_{i\bot})=(e_i-k_{iz},e_i+k_{iz},k_{i\bot})=
  (\frac{m_i^2+k_{i\bot}^2}{x_iM_0},x_iM_0,k_{i\bot}).
 \eeq
It is easy to obtain
 \begin{eqnarray}
  M_0&=&e_1+e_2, \non\\
  e_i&=&\frac{x_iM_0}{2}+\frac{m_i^2+k_{i\perp}^2}{2x_iM_0}
      =\sqrt{m_i^2+k_{i\bot}^2+k_{iz}^2},\non\\
 k_{iz}&=&\frac{x_iM_0}{2}-\frac{m_i^2+k_{i\perp}^2}{2x_iM_0},
 \end{eqnarray}
where $e_i$ denotes the energy of the $i$-th constituent. The
momenta $k_{i\bot}$ and $k_{iz}$ constitute a momentum vector
$\vec k_i=(k_{i\bot}, k_{iz})$ and correspond to the components in
the transverse and $z$ directions, respectively.  $ m_1 (m_2)$ is
the mass of the heavy quark (the light diquark).

The momentum-space function
$\Psi_{\Sigma_c}^{SS_z}(\tilde{p}_1,\tilde{p}_2,\lambda_1,\sigma)$
in Eq. (\ref{eq:lfbaryon}) is expressed as
\begin{eqnarray*}
 \Psi_{\Sigma_c}^{SS_z}(\tilde{p}_1,\tilde{p}_2,\lambda_1,\sigma)=&&
  \left\la\lambda_1\left|\mathcal{R}^{\dagger}_M(x_1,k_{1\perp},m_1)
   \right|s_1\right\ra
 \left\la
\frac{1}{2}s_1 ;1 \sigma\left|\frac{1}{2}S_z\right\ra
 \varphi(x,k_{\perp})\right.,
\end{eqnarray*}
where $ \langle\frac{1}{2}s_1;1 s_2|\frac{1}{2}S_z\rangle$ is the
C-G coefficients corresponding to the concerned transition and
$s_1 (m)$ are the spin projections of the constituents the heavy
quark (diquark).

\begin{eqnarray}\label{CG}
\langle\frac{1}{2}s_1;1
\sigma|\frac{1}{2}S_z\rangle=A_1\bar{u}(p_1,s_1)\frac{-\gamma_5\varepsilon^*\!\!\!\!\!\slash(\bar
P,\sigma)}{\sqrt{3}} u(\bar {P},S_z).
\end{eqnarray}

Calculating the modular squares of the
two sides and summing over all the polarizations, one
obtains
\begin{eqnarray}
A_1=\frac{1}{\sqrt{2(M_0m_1+p_1\cdot\bar{P})}}.
\end{eqnarray}

$A_1$ also can be obtained by substituting the explicit expressions of $\bar{u}(p_1,s_1)$
$\varepsilon^*\!\!\!\!\!\slash(\bar P,\sigma)$ $u(\bar {P},S_z)$
into Eq.(2), where $S_z$ is also
spin projection of baryon\cite{pentaquark1,pentaquark2}.

A Melosh transformation brings the the matrix elements from the
spin-projection-on-fixed-axis representation into the helicity
representation and  is explicitly written as
$$\left\la\lambda_1\left|\mathcal{R}^{\dagger}_M(x_1,k_{1\perp},m_1)
   \right|s_1\right\ra=\frac{\bar{u}(k_1,\lambda_1)u(k_1,s_1)}{2m_1}.$$

Following Refs. \cite{pentaquark1,pentaquark2}, the
Melosh-transformed matrix is expressed as
\begin{eqnarray}
 &&\left\la\lambda_1\left|\mathcal{R}^{\dagger}_M(x_1,k_{1\perp},m_1)
   \right|s_1\right\ra \left\la \frac{1}{2}s_1;1 \sigma\left|
   \frac{1}{2}S_z\right\ra\right.\nonumber\\&&=\frac{1}{\sqrt{6(p_1\cdot
   \bar P+m_1M_0)}}\bar{u}(p_1,\lambda_1)[-\gamma_5
\varepsilon^*\!\!\!\!\!\slash(\bar P,\sigma)] u(\bar {P},S_z),
\end{eqnarray}
where
\begin{eqnarray}\label{polar}
 &&\varepsilon^\mu(\bar P,0)=(\bar P^+,\frac{\bar
P^2_\bot-M_0^2}{\bar
P^+},P_\bot),\nonumber\\&&\varepsilon^\mu(\bar
P,\pm1)=(0,\frac{2}{\bar P^+}\varepsilon_\perp(\pm1)\cdot \bar
P_\bot,\varepsilon_\perp(\pm1)),\nonumber\\&&\varepsilon_\perp(\pm1)=\mp(1,\pm
i)/\sqrt{2}
\end{eqnarray}
with
$\varphi(x,k_\perp)=4(\frac{\pi}{\beta^2})^{3/4}\frac{e_1e_2}{x_1x_2M_0}{\rm
exp}(\frac{-\mathbf{k}^2}{2\beta^2})$.
In this way, the finial expression
\begin{eqnarray}\label{psi1}
\Psi_{\Sigma_c}^{SS_z}(\tilde{p}_1,\tilde{p}_2,\lambda_1,\sigma)=\frac{1}{\sqrt{6(p_1\cdot
   \bar P+m_1M_0)}}\bar{u}(p_1,\lambda_1)[-\gamma_5
\varepsilon^*\!\!\!\!\!\slash(\bar P,\sigma)] u(\bar {P},S_z)
 \varphi(x,k_{\perp}),
\end{eqnarray}
which is the same as that given in those
papers\cite{pentaquark1,pentaquark2}.

 For the baryon $\Sigma^*_Q$ with the
total spin $S=3/2$, the wavefunction is a bit more complicated as
\begin{eqnarray}
 \Psi^{SS_z}(\tilde{p}_1,\tilde{p}_2,\lambda_1,\sigma)=&&
  \left\la\lambda_1\left|\mathcal{R}^{\dagger}_M(x_1,k_{1\perp},m_1)
   \right|s_1\right\ra
 \left\la
\frac{1}{2}s_1 ;1 \sigma\left|\frac{3}{2}S_z\right\ra
 \varphi(x,k_{\perp})\right.,
\end{eqnarray}
and one has
\begin{eqnarray}
\langle\frac{1}{2}s_1;1
\sigma|\frac{3}{2}S_z\rangle=A'_1\bar{u}(p_1,s_1)\varepsilon^{*\alpha}(\bar
P,\sigma) u_\alpha(\bar {P},S_z).
\end{eqnarray}

Evaluating the modular squares of the
left and right sides separatively and summing over all the
polarizations one has
\begin{eqnarray}
A'_1={\frac{1}{\sqrt{2(p_1\cdot
   \bar P+m_1M_0)}}}.
\end{eqnarray}

The finial expression is
\begin{eqnarray}\label{psi2}
 \Psi_{\Sigma^*_c}^{SS_z}(\tilde{p}_1,\tilde{p}_2,\lambda_1,\sigma)=&&\frac{1}{\sqrt{2(p_1\cdot
   \bar P+m_1M_0)}}\bar{u}(p_1,s_1)\varepsilon^{*\alpha}(\bar P,\sigma )
u_\alpha(\bar {P},S_z)
 \varphi(x,k_{\perp}).
\end{eqnarray}

All other relevant notations can be found in
Ref.\cite{pentaquark2}.

\subsection{the vertex functions of $\Sigma_{Q}$ and
$\Sigma^*_{Q}$ in Scheme II}

For the axial vector diquark inside the baryons $\Sigma_c$ or
$\Sigma^*_c$, a natural conjecture is that the polarization vector
$\varepsilon$ in the finial expressions of $\Psi_{\Sigma_c}$ and $
\Psi_{\Sigma^*_c}$ should depend on the momentum of the diquark:
$p_2$, instead of the total momentum $\bar P$. With this
consideration,  we rewrite the expressions of $\Psi_{\Sigma_c}$
and $ \Psi_{\Sigma^*_c}$ as
\begin{eqnarray}\label{psi3}
&&\Psi_{\Sigma_c}^{SS_z}(\tilde{p}_1,\tilde{p}_2,\lambda_1,\sigma)=\frac{A_2}{\sqrt{6(p_1\cdot
   \bar P+m_1M_0)}}\bar{u}(p_1,\lambda_1)[-\gamma_5
\varepsilon^* \!\!\!\!\!\slash (p_2,\sigma)] u(\bar {P},S_z)
 \varphi(x,k_{\perp}),\\&&\Psi_{\Sigma^*_c}^{SS_z}(\tilde{p}_1,\tilde{p}_2,\lambda_1,\sigma)=\frac{A_2'}{\sqrt{2(p_1\cdot
   \bar P+m_1M_0)}}\bar{u}(p_1,s_1)\varepsilon^{*\alpha}(p_2,\sigma )
u_\alpha(\bar {P},S_z)
 \varphi(x,k_{\perp}),
\end{eqnarray}
with
$A_2=\sqrt{\frac{12(M_0m_1+p_1\cdot\bar{P})}{12M_0m_1+4p_1\bar{P}+8p_1\cdot
p_2 p_2\cdot \bar{P}/m_2^2}}$ and
$A_2'=\sqrt{\frac{3M_0^2m_2^2}{2M_0^2m_2^2+(\bar{P}\cdot p_2)^2}}$
which can be obtained by normalizing the state
$|\Sigma_Q(P,S,S_z)\rangle$ as, \beq\label{A12}
 \la
 \Sigma_Q(P',S',S'_z)|\Sigma_Q(P,S,S_z)\ra=2(2\pi)^3P^+
  \delta^3(\tilde{P}'-\tilde{P})\delta_{S'S}\delta_{S'_zS_z}.
 \eeq

In fact, the coefficients $A_2$ and $A_2'$ would change to 1 when
$p_2$ and $m_2$ are replaced by $\bar P$ and $M_0$ respectively,
thus the expressions  in Eq. (\ref{psi1}) and (\ref{psi2}) are
recovered. In Ref. \cite{Chua:2018lfa} the authors present a
similar vertex function for $\Sigma_Q$ which adds  a term
$\varepsilon^* (p_2,\sigma)\cdot \bar P$ into Eq. (18) of this work. Since
a normalization condition is required for every vertex function
the difference would not seriously affect the results.
\subsection{Calculating the  form factors of $\Sigma_b\to\Sigma_c^*$ in LFQM}

\begin{figure}
\begin{center}
\scalebox{0.8}{\includegraphics{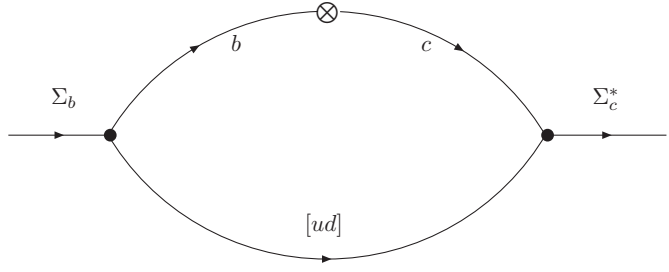}}
\end{center}
\caption{The Feynman diagram for $\Sigma_{b}\to\Sigma_{c}^*$
transitions, where $\bigotimes$ denotes $V-A$ current
vertex.}\label{t1}
\end{figure}

The leading order Feynman diagram responsible for the
$\Sigma_{b}\to\Sigma_{c}^*$ weak decay is shown in Fig. \ref{t1}.
Following the approach given in Ref.\cite{pentaquark1,pentaquark2}
the transition matrix element can be computed with the
wavefunctions of $\mid \Sigma_{b}(P,S,S_z) \ra$ and $\mid
\Sigma_{c}^*(P,S,S_z) \ra$ in the Scheme I,
\begin{eqnarray}\label{s00}
&& \la \Sigma_{c}^*(P',S_z') \mid \bar{c}
\gamma^{\mu} (1-\gamma_{5}) b \mid \Sigma_{b}(P,S_z) \ra  \nonumber \\
 &=& \int\{d^3 \tilde p_2\}\frac{\phi_{\Sigma_{c}^*}^*(x',k'_{\perp})
  \phi_{\Sigma_b}(x,k_{\perp})}{2\sqrt{3p^+_1p'^+_1(p_1\cdot \bar{P}+m_1M_0)
 (p'_1\cdot \bar{P'}+m'_1M'_0)}}\nonumber \\
  &&\times  \bar{u}_\alpha(\bar{P'},S'_z)[\varepsilon^{\alpha}(\bar P',\sigma')]
  (p_1\!\!\!\!\!\slash'+m'_1)\gamma^{\mu}(1-\gamma_{5})
  (p_1\!\!\!\!\!\slash+m_1) [-\gamma_5\varepsilon^*\!\!\!\!\!\slash(\bar P,\sigma)] u(\bar{P},S_z),
\end{eqnarray}
where
 \beq
m_1=m_b, \qquad m'_1=m_c, \qquad m_2=m_{[ud]},
 \eeq
and $Q$ ($Q'$) represents the heavy quark $b$ ($c$), $p_1$
($p'_1$) denotes the four-momentum of the heavy quark $b$ ($c$), $P$
($P'$) stands as the four-momentum of $\Sigma_{b}$ ($\Sigma_{c}^*$).
Setting $\tilde{p}_2=\tilde{p}'_2$, we have
 \beq
 x'=\frac{P^+}{P^{'+}}x, \qquad \qquad
 k'_{\perp}=k_{\perp}+x_2q_{\perp}.
 \eeq

Instead, with the Scheme II the transition matrix element is
obtained by replacing $\varepsilon^*(\bar P',\sigma)$ and
$\varepsilon(\bar P,\sigma)$ by $A_2'\varepsilon^*(p_2,\sigma)$
and $A_2\varepsilon(p_2,\sigma)$ in Eq. (\ref{s00}). It is noted
that one can use the formula
$\varepsilon^*(p_2,\sigma)^\mu\varepsilon(p_2,\sigma)^\nu=-g_{\mu\nu}+\frac{p_2^\mu
p_2^\nu}{m_2^2}$ to deal with $\varepsilon^*(p_2,\sigma)$ but he
needs the components of the polarization vectors of
$\varepsilon^*(\bar P',\sigma)$ and $\varepsilon(\bar P,\sigma)$
while summing over the polarizations. The transition matrix element
is calculated in the $q^+=0$ reference frame.

The form factors for the weak transition $\Sigma_b\rightarrow
\Sigma_{c}^*$  are defined in the standard way as
\begin{eqnarray}\label{s1}
&& \la \Sigma_{c}^*(P',S',S_z') \mid \bar{c}\gamma^{\mu}
 (1-\gamma_{5})b \mid \Sigma_{b}(P,S,S_z) \ra  \non \\
 &&= \bar{u}_{\alpha}(P',S'_z) \left[ \gamma^{\mu} P^\alpha \frac{f_{1}(q^{2})}{M_{\Sigma_b}}
 + \frac{f_{2}(q^{2})}{M_{\Sigma_b}M_{\Sigma_b}} P^\alpha P^\mu+
 \frac{f_{3}(q^{2})}{M_{\Sigma_b}M_{\Sigma^*_c}} P^\alpha P'^\mu+f_4(q^{2})g^{\alpha \mu}
 \right] u(P,S_z)- \nonumber \\
 &&\bar u_{\alpha}(P',S'_z)\left[\gamma^{\mu} P^\alpha \frac{g_{1}(q^{2})}{M_{\Sigma_b}}
 + \frac{g_{2}(q^{2})}{M_{\Sigma_b}M_{\Sigma_b}} P^\alpha P^\mu+
 \frac{g_{3}(q^{2})}{M_{\Sigma_b}M_{\Sigma^*_c}} P^\alpha P'^\mu+g_4(q^{2})g^{\alpha \mu}
 \right]\gamma_{5} u(P,S_z),
\end{eqnarray}
where  $q \equiv P-P'$, $Q$ and $Q'$ denote $b$ and $c$,
respectively.

The baryon spinors $u(P,S_z)$ and $\bar u(P',S_z)$ which appear in
the above formula are different from $ u(\bar P,S_z)$ and $\bar
u(\bar P',S_z)$. The  spinors in Eq. (\ref{s00}) which are
functions of momenta $\bar P$ and $\bar P'$ do not correspond to
on-mass-shell states, whereas $ u( P,S_z)$ and $\bar u( P',S_z)$
stand for the the on-shell baryons. Concretely, $\bar
P^{(')2}+M^2_{\Sigma_b(M_{\Sigma_c^*})}\neq
E^2_{\Sigma_b(M_{\Sigma_c^*})}$, as $\bar P$ and $\bar P'$ are the
sums of the momenta of the involved constituents (quark and
diquark) which are on their mass shells. Thus in principle one
cannot obtain the form factors $f_1$, $f_2$, $f_3$, $f_4$, $g_1$,
$g_2$, $g_3$ and $g_4$  from Eq.(\ref{s00}) at all.  In Eq.
(\ref{s00}), the hadronic matrix elements are calculated in the
light-front-quark model in the un-physical regions. Since $P$ and
$P'$ obey the relations $E^{(')2}=P^{(')2}+M^{(')2}$, the form
factors in Eq. (\ref{s1}) are not directly extracted from Eq.
(\ref{s00}). To remedy the conflict, if it is assumed that the
form factors are the same in both physical and unphysical regions,
we can extrapolate Eq. (\ref{s1}) to the following equation (Eq.
(\ref{s2})) where the spinors on the right side are off-shell,
with the same form factors.

The Eq. (\ref{s1}) is re-written as
\begin{eqnarray}\label{s2}
&& \la \Sigma_{c}^*( P',S',S_z') \mid \bar{c}\gamma_{\mu}
 (1-\gamma_{5})b \mid \Sigma_{b}( P,S,S_z) \ra  \non \\
 &&=\bar{u}_{\alpha}(\bar P',S'_z) \left[ \gamma^{\mu}\bar P^\alpha \frac{f_{1}(q^{2})}{M_{\Sigma_b}}
 + \frac{f_{2}(q^{2})}{M_{\Sigma_b}M_{\Sigma_b}}\bar P^\alpha \bar P^\mu+\frac{f_{3}(q^{2})}{M_{\Sigma_b}M_{\Sigma^*_c}}\bar P^\alpha \bar P'^\mu
 +f_4(q^{2})g^{\alpha \mu}
 \right] u(\bar P,S_z) -\nonumber \\
 &&\bar u_{\alpha}(\bar P',S'_z)\left[\gamma^{\mu}\bar P^\alpha \frac{g_{1}(q^{2})}{M_{\Sigma_b}}
 + \frac{g_{2}(q^{2})}{M_{\Sigma_b}M_{\Sigma_b}}\bar P^\alpha \bar P^\mu+\frac{g_{3}(q^{2})}
 {M_{\Sigma_b}M_{\Sigma^*_c}}\bar P^\alpha \bar P'^\mu+g_4(q^{2})g^{\alpha \mu}
 \right]\gamma_{5} u(\bar P,S_z),
\end{eqnarray}
here $f_1$, $f_2$, $f_3$, $f_4$, $g_1$, $g_2$, $g_3$ and $g_4$ are
supposed to be invariant for any definite $q^2$ no matter $q=P'-P$ or $q=\bar P'-\bar P$.

Multiplying the following expressions $\bar u(\bar P,S_z)\gamma^{\mu}\bar P^\beta
{u}_{\beta}(\bar P',S'_z)$, $\bar u(\bar P,S_z)\bar P'^{\mu}\bar
P^\beta{u}_{\beta}(\bar P',S'_z)$ , $\bar u(\bar P,S_z) \bar
P^{\mu}\bar P^\beta{u}_{\beta}(\bar P',S'_z)$, $\bar u(\bar P,S_z)
g^{\mu \beta}{u}_{\beta}(\bar P',S'_z)$, $ \bar u(\bar
P,S_z)\gamma^{\mu}\bar P^\beta \gamma_{5}{u}_{\beta}(\bar
P',S'_z)$, $\bar u(\bar P,S_z)\bar P'^{\mu}\bar
P^\beta\gamma_{5}{u}_{\beta}(\bar P',S'_z)$ , $\bar u(\bar P,S_z)
\bar P^{\mu}\bar P^\beta$ $\gamma_{5}{u}_{\beta}(\bar P',S'_z)$,
and $\bar u(\bar P,S_z) g^{\mu \beta}\gamma_{5}{u}_{\beta}(\bar
P',S'_z)$ to the right sides of Eq.(\ref{s00}) and
Eq.(\ref{s2}) ($\mu=+$), several algebraic equations are obtained.
Then by solving them we achieve $f_1$, $f_2$, $f_3$, $f_4$, $g_1$,
$g_2$, $g_3$ and $g_4$ (See Appendix for detail).

\subsection{ The generalized Isgur-Wise functions for the transition}

Under the heavy quark limit ($m_Q\to\infty$)\cite{HQS},  the eight
form factors $f_i,~g_i$ (i=1,2,3,4) are no longer independent and
the matrix elements are totally determined by two Isgur-Wise
functions $\xi_1(v\cdot v')$ and $\xi_2(v\cdot v')$ and they are
defined through the following
expression\cite{Korner:1992uw,Ebert:2006rp}
  \begin{eqnarray}\label{s10}
  && < \Sigma^*_{c}(v',S_z')\mid\bar{Q}_{v'}' \gamma_{\mu}
   (1-\gamma_{5}) Q_v \mid \Sigma_{b}(v,S_z)>\nonumber\\&& = \frac{1}{\sqrt{3}} [g^{\alpha\beta}\xi_1(\omega)-v^\alpha v'^\beta\xi_2(\omega)]
  \bar{u}_\alpha(v',S'_z)\gamma^{\mu}(1-\gamma_{5})(\gamma_\beta+v_\beta)\gamma_5
  u(v,S_z),
  \end{eqnarray}
where $\omega\equiv v\cdot v'$,
and
 \begin{eqnarray}\label{IWF}
&&f_1=-\frac{\xi_1-\xi_2-\xi_2w}{\sqrt{3}}  , f_2=0 ,
f_3=-\frac{2\xi_2}{\sqrt{3}} ,
f_4=\frac{2\xi_1}{\sqrt{3}}\nonumber\\&&g_1=-\frac{\xi_1+\xi_2-\xi_2w}{\sqrt{3}}
, g_2=0 , g_3=\frac{2\xi_2}{\sqrt{3}} ,
g_4=-\frac{2\xi_1}{\sqrt{3}}.
 \end{eqnarray}

Thus using these relations one can immediately
obtain the form factors with the heavy quark limit from the
Isgur-Wise functions.


With the re-normalized wavefunctions\cite{pentaquark1,pentaquark2}
 \beq
 \mid\Sigma_{Q}(P,S_z)\ra &\to& \sqrt{M_{\Sigma_Q}}
 \mid\Sigma_{Q}(v,S_z)\ra,
  \non \\
 u(\bar{P},S_z) &\to& \sqrt{m_Q}u(v,S_z) \non \\
 \phi_{\Sigma_Q}(x,k_{\bot})&\to &\sqrt{\frac{m_Q}{X}}\Phi(X,k_{\bot}),
 \eeq
and
\begin{eqnarray}\label{s12}
 &&M_{\Sigma_Q}\to m_Q, \qquad ~~ M_0\to m_Q,\non\\
 &&e_1\to m_Q, \qquad\qquad \non\\
 &&e_2\to v\cdot p_2=\frac{m_2^2+k_{\perp}^2+X^2}{2X},\non\\
 &&\vec k^2\to (v\cdot p_2)^2-m_2^2,\non\\
 &&p_1\!\!\!\!\!\slash+m_1\to m_Q(v\!\!\!\slash+1)\non \\
 &&\frac{e_1e_2}{x_1x_2M_0}\to\frac{m_Q}{X}(v\cdot p_2),
\end{eqnarray}
the transition matrix elements obtained in the previous section are
re-formulated under the heavy quark limit.

In the Scheme I
\begin{eqnarray}\label{sw14}
 &&\la\Sigma_{c}^*(v',S_z') \mid \bar{c}_{v'}
   \gamma^{\mu}(1-\gamma_{5})b_v\mid\Sigma_{b}(v,S_z) \ra  \non\\
 &=& -\int{\frac{dX}{X}\frac{d^2k_{\perp}}{2(2\pi)^3}}
     {\Phi(X,k_{\perp})\Phi(X',k_{\perp}^{\prime})}
    \frac{1}{\sqrt{3}} \bar{u}_\alpha(v',S'_z)\gamma^\mu(1-\gamma_5) (\gamma_\beta+v_\beta)\gamma_5\nonumber\\&&
   u(v,S_z)\varepsilon^{\alpha*}(\bar P',\sigma)\varepsilon^{\beta}(\bar P,\sigma),
\end{eqnarray}
with
 \beq
 \Phi(X,k_{\bot})=4\sqrt{v\cdot p_2}\left(\frac{\pi}{\beta^2_\infty}
  \right)^{\frac{3}{4}}{\rm exp}\left(-\frac{(v\cdot p_2)^2-m^2_2}
  {2\beta^2_\infty}\right),\nonumber\\
  \Phi(X',k'_{\bot})=4\sqrt{v'\cdot p_2}\left(\frac{\pi}{\beta^2_\infty}
  \right)^{\frac{3}{4}}{\rm exp}\left(-\frac{(v'\cdot p_2)^2-m^2_2}
  {2\beta^2_\infty}\right),
 \eeq
where $\beta_{\infty}$ denotes the value of $\beta$ in the heavy
quark limit.

The transition matrix element is
          \begin{eqnarray}\label{s141}
        &&<\Sigma^*_{c}(v',S_z') \mid \bar{c}_{v'}
          \gamma^{\mu}(1-\gamma_{5})b_v\mid\Sigma_{b}(v,S_z) >
       \nonumber\\&& =- \int{\frac{dX}{X}\frac{d^2k_{\perp}}{2(2\pi)^3}}
        {\Phi(X,k_{\perp})\Phi(X',k_{\perp}^{\prime})} \frac{1}{\sqrt{3}} \bar{u_\alpha}(v',S'_z)
       \gamma^\mu(1-\gamma_5) (\gamma_\beta+v_\beta)\gamma_5 \nonumber\\&&
      u(v,S_z)(
       a_1 g^{\alpha\beta}+a_2 v^\alpha v'^\beta+a_3 v'^\alpha v^\beta+a_4 v'^\alpha v'^\beta+a_5 v^\alpha v^\beta).
       \end{eqnarray}
Using the relation
$\bar{u}'\gamma_5(v'\!\!\!\!\slash+1)=(v\!\!\!\slash+1)\gamma_5u=0$,
the terms involving  $a_3$, $a_4$ and $a_5$ do not contribute to the
transition, thus
 \begin{eqnarray}\label{s142}
        &&<\Sigma^*_{c}(v',S_z') \mid \bar{c}_{v'}
          \gamma^{\mu}(1-\gamma_{5})b_v\mid\Sigma_{b}(v,S_z) >
       \nonumber\\&& =- \int{\frac{dX}{X}\frac{d^2k_{\perp}}{2(2\pi)^3}}
        {\Phi(X,k_{\perp})\Phi(X',k_{\perp}^{\prime})} \frac{1}{\sqrt{3}} \bar{u_\alpha}(v',S'_z)
       \gamma^\mu(1-\gamma_5) (\gamma_\beta+v_\beta)\gamma_5\nonumber\\&&
       u(v,S_z)(
       a_1 g^{\alpha\beta}+a_2 v^\alpha v'^\beta),
       \end{eqnarray}
and
       \begin{eqnarray}
          && a_1=-1 \nonumber\\&&
           a_2=\frac{1}{\omega+1}
       \end{eqnarray}

 In the Scheme II
\begin{eqnarray}\label{sw14}
 &&\la\Sigma_{c}^*(v',S_z') \mid \bar{c}_{v'}
   \gamma^{\mu}(1-\gamma_{5})b_v\mid\Sigma_{b}(v,S_z) \ra  \non\\
 &=& -\int{\frac{dX}{X}\frac{d^2k_{\perp}}{2(2\pi)^3}}
     {\Phi(X,k_{\perp})\Phi(X',k_{\perp}^{\prime})}
    \frac{1}{\sqrt{3}} \bar{u}_\alpha(v',S'_z)\gamma^\mu(1-\gamma_5) (\gamma_\beta+v_\beta)\gamma_5\nonumber\\&&
   u(v,S_z)(\frac{p_2^\alpha p_2^\beta}{m_2^2}-g^{\alpha\beta}),
\end{eqnarray}
with
 \beq
 \Phi(X,k_{\bot})=\sqrt{\frac{24}{16+8v\cdot
p_2^2/m_2^2}}4\sqrt{v\cdot p_2}\left(\frac{\pi}{\beta^2_\infty}
  \right)^{\frac{3}{4}}{\rm exp}\left(-\frac{(v\cdot p_2)^2-m^2_2}
  {2\beta^2_\infty}\right),\nonumber\\
  \Phi(X',k'_{\bot})=\sqrt{\frac{3}{2+v'\cdot
p_2^2/m_2^2}}4\sqrt{v'\cdot p_2}\left(\frac{\pi}{\beta^2_\infty}
  \right)^{\frac{3}{4}}{\rm exp}\left(-\frac{(v'\cdot p_2)^2-m^2_2}
  {2\beta^2_\infty}\right),
 \eeq
and
       \begin{eqnarray}\label{a1a2II}
          && a_1=-\frac{(w^2-1)p_2^2+2v\cdot p_2 v'\cdot p_2 \omega-(v'\cdot p_2)^2
           -(v\cdot p_2)^2 }{2m_2^2(\omega^2-1)}, \nonumber\\&&
           a_2=-\frac{\omega(\omega^2-1)p_2^2-2v\cdot p_2 v'\cdot p_2(2\omega^2+1)
           +3\omega[(v'\cdot p_2)^2+(v\cdot
           p_2)^2]}{2m_2^2(\omega^2-1)^2}.
       \end{eqnarray}

Comparing Eq.(\ref{s141}) with Eq. (\ref{s10}), we get
       \begin{eqnarray}\label{s143}
       && \xi_1 = -\int{\frac{dX}{X}\frac{d^2k_{\perp}}{2(2\pi)^3}}
        {\Phi(X,k_{\perp})\Phi(X',k_{\perp}^{\prime})}a_1,
       \end{eqnarray}
        \begin{eqnarray}\label{s144}
        && \xi_2 = \int{\frac{dX}{X}\frac{d^2k_{\perp}}{2(2\pi)^3}}
         {\Phi(X,k_{\perp})\Phi(X',k_{\perp}^{\prime})}a_2.
        \end{eqnarray}
Here $\xi_1$ and $\xi_2$ are similar to the forms appearing in
Eq.(4.18) and Eq. (4.19) of Ref.\cite{pentaquark2}. We are able to
directly evaluate them in the time-like region by choosing a
reference frame with $q_\perp=0$. It is noted that the components
of $\varepsilon(\bar P',\sigma)$ and $\bar P'$ are different from
those in the reference frame with $q^+=0$.

\section{Numerical Results}

In order to evaluate the transition rate of $\Sigma_b\to\Sigma^*_c$
one needs to pre-set all input parameters. The baryon masses
$\Mb=5.811$ GeV, $\Mc=2.517$ GeV are taken from\cite{PDG10}. The
heavy quark masses $m_b$ and $m_c$ are set following
Ref.\cite{Cheng:2003sm}. In our calculation, the mass of the light
axial vector diquark $m_{[ud]_{_V}}$ is set to be 770
MeV\cite{Korner:1992uw}. It can be conjectured that the diquark mass
$m_{[ud]_{_V}}$ is close to the mass of an $s$ quark (here we
consider the constituent quark masses instead of current quark
masses), thus we set $\beta_{b[ud]}=\beta_{b\bar s}$ and
$\beta_{c[ud]}=\beta_{c\bar s}$ while the values of the
corresponding model parameters are adopted from the meson
cases\cite{Cheng:2003sm}. The relevant input parameters are
collected in Table \ref{Tab:t1}.

First of all, we need to numerically calculate the form factors,
then using them we are able to predict the rates of semi-leptonic
processes $\Sigma_b\rightarrow \Sigma^*_c l\bar{\nu}_l$ and
non-leptonic decays $\Sigma_b\to \Sigma^*_c M^-$ ($M$ represents
$\pi,~ K,~ \rho,~ K^*,~ a_1$ etc.).

\begin{table}
\caption{Quark mass and the parameter $\beta$ (in units of
 GeV).}\label{Tab:t1}
\begin{ruledtabular}
\begin{tabular}{ccccc}
  $m_c$  & $m_b$  &$m_{[ud]}$ & $\beta_{c[ud]}$ & $\beta_{b[ud]}$ \\\hline
  $1.3$  & $4.4$  & 0.77     & $0.45$         & 0.50
\end{tabular}
\end{ruledtabular}
\end{table}

\subsection{$\Sigma_b\to \Sigma^*_c$ form factors and the Isgur-Wise
functions in Scheme I}

\begin{figure}
\begin{center}
\scalebox{0.8}{\includegraphics{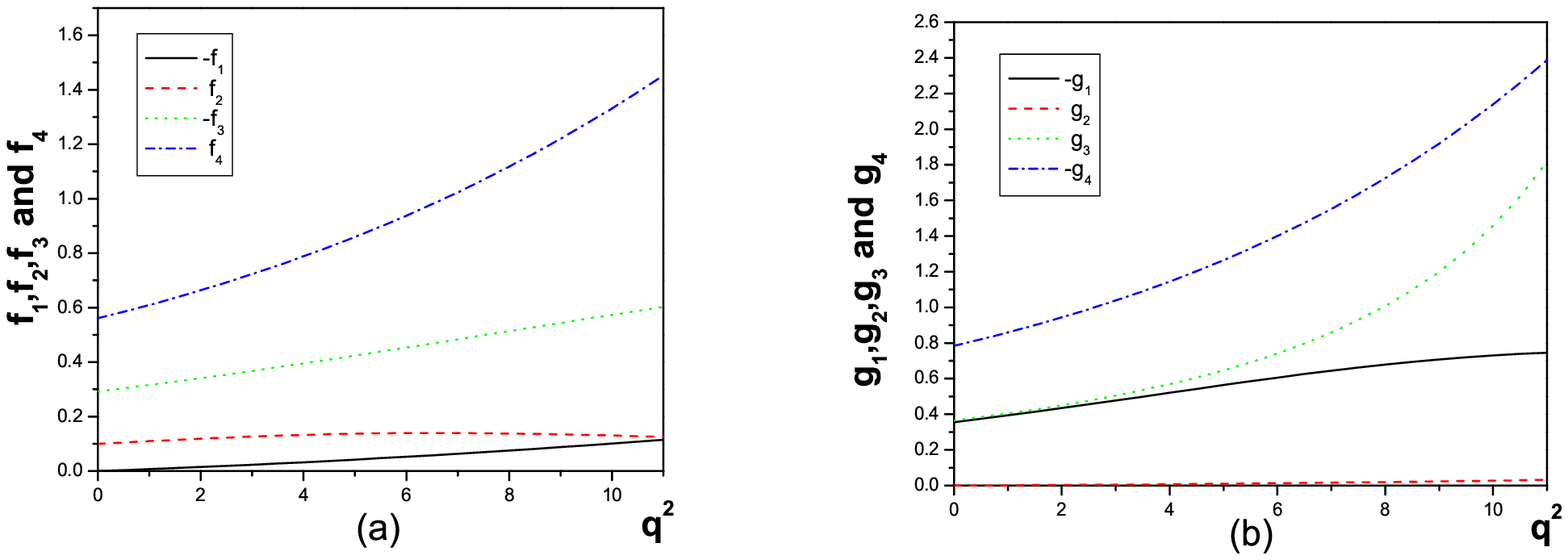}}
\end{center}
\caption{(a)  the form factors  $f_i\; (i=1,2,3,4)$and (b) the
form factors $g_i\; (i=1,2,3,4)$ } in Scheme I\label{f21}
\end{figure}

Since these  form factors $f_i\; (i=1,2,3,4)$ and $g_i\;
(i=1,2,3,4)$ are evaluated in the frame $q^+=0$ i.e.
$q^2=-q^2_{\perp}\leq 0$ (the space-like region) one needs to
extend them into the time-like region. As commonly adopted in
literature, we may employ a three-parameter form
factor\cite{pentaquark2}
 \begin{eqnarray}\label{s145}
 F(q^2)=\frac{F(0)}{\left(1-\frac{q^2}{M_{\Sigma_b}^2}\right)
  \left[1-a\left(\frac{q^2}{M_{\Sigma_b}^2}\right)
  +b\left(\frac{q^2}{M_{\Sigma_b}^2}\right)^2\right]},
 \end{eqnarray}
where $F(q^2)$ denotes the form factors $f_i\; (i=1,2,3,4)$ and
$g_i\; (i=1,2,3,4)$. However, this extrapolation shown in Eq.
(\ref{s145}) suffers a fatal disadvantage that if  $F(0)=0$, the
form factor would remain zero for $q^2\neq 0$, which obviously is
invalid. Therefore we employ an alternative three-parameter
extrapolation as
 \begin{eqnarray}\label{s145p}
 F(q^2)=F(0)-a\frac{q^2}{M_{\Sigma_b}^2}+b\left(\frac{q^2}{M_{\Sigma_b}^2}\right)^2.
 \end{eqnarray}
Using the form factors in the space-like region we may calculate
numerically the parameters $a,~b$ and $F(0)$, namely fixing
$F_i(q^2\leq 0)$. As discussed in previous section, these form
factors are extended into the physical region with $q^2\geq 0$
through Eq. (\ref{s145}) or (\ref{s145p}). The fitted values of
$a,~b$ and $F(0)$ in the form factors $f_i\; (i=1,2,3,4)$ and
$g_i\; (i=1,2,3,4)$ are presented in Table \ref{Tab:t21}. The
dependence of the form factors on $q^2$ is depicted in Fig.
\ref{f21}.
\begin{table}
\caption{The $\Sigma_b\to \Sigma^*_c$ form factors given in the
  three-parameter form in Scheme I.}\label{Tab:t21}
\begin{ruledtabular}
\begin{tabular}{cccc}
  $F$    &  $F(0)$ &  $a$  &  $b$ \\\hline
  $f_1$  &   0     &  0.225    & -0.388   \\
$f_2$  &   0.100     &   2.23    & 8.57   \\
  $f_3$  &    -0.292   &   1.66    &  2.42  \\
  $f_4$  &     0.561   &   1.80    &  1.50  \\
  $g_1$  &      -0.356    &     2.47 &  4.80  \\
  $g_2$  &      0    &    0.044  &  -0.167  \\
    $g_3$  &      0.365   &     2.38  & 0.69\\
  $g_4$  &      -0.784   &     2.07  & 1.49
\end{tabular}
\end{ruledtabular}
\end{table}

Now let us turn to calculate the form factors in the HQET. In the
heavy quark limit, $\beta^\infty=0.50$ GeV  is used for $\Sigma_b$
and $\Sigma_c^*$.
The Isgur-Wise function is  parameterized as
 \begin{eqnarray}
 \xi(\omega)=1-\rho^2(\omega-1)+\frac{\sigma^2}{2}(\omega-1)^2,
 \end{eqnarray}
where $\rho^2\equiv-\frac{d\xi(\omega)}{d\omega}|_{\omega=1}$ is
the slope parameter and
$\sigma^2\equiv\frac{d^2\xi(\omega)}{d\omega^2}|_{\omega=1}$ is
the curvature of the Isgur-Wise function. To fit some available data, we write up
the expressions with definite values as
 \beq
\xi_1=1-1.90(\omega-1)+1.58(\omega-1)^2\\
\xi_2=0.50[1-2.36(\omega-1)+2.28(\omega-1)^2].
 \eeq

\begin{figure}[hhh]
\begin{center}
\scalebox{0.8}{\includegraphics{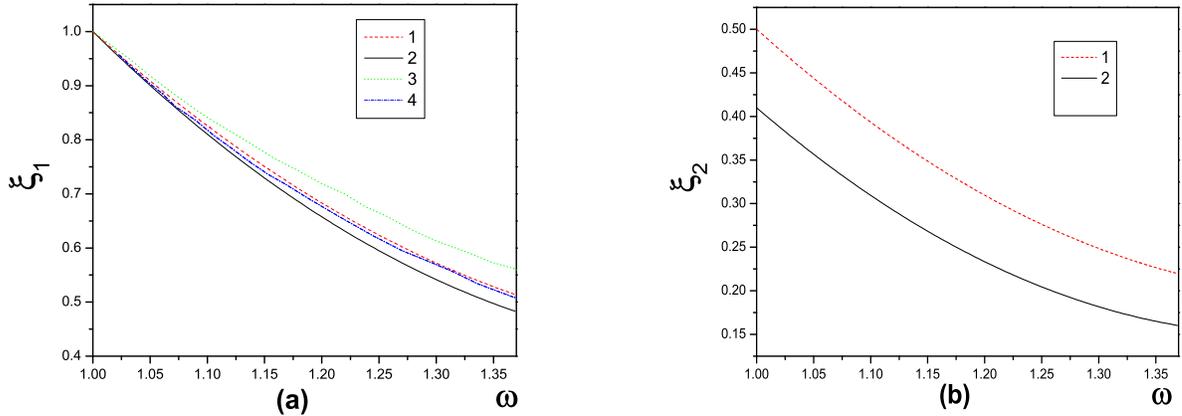}}
\end{center}
\caption{(a) The Isgur-Wise function $\xi_1(\omega)$ for $\Sigma_{b}
\rightarrow \Sigma^*_{c}$ where the line 1  and 2 are our results of
$\xi_1(\omega)$ in Scheme I and II respectively, whereas the line 3
and 4  are given in Ref.\cite{Ivanov:1996fj} (b) The Isgur-Wise
function $\xi_2(\omega)$ for $\Sigma_{b} \rightarrow \Sigma^*_{c}$
where the line 1  and 2 also are the results in Scheme I and II
respectively.}\label{f3}
\end{figure}

The dependence of the Isgur-Wise function $\xi_1$ on $\omega$  is
depicted in Fig.\ref{f3}(a).  Let us compare $\xi_1$ obtained in
this work with that given in Ref.\cite{Ivanov:1996fj}. One can
notice that $\xi_1|_{\omega=1}=1$ holds, which is the mandatory
normalization of the Isgur-Wise function. The dashed line (line 1)
and the solid line (line 2) are our results in Scheme I and Scheme
II respectively with the diquark mass being 770 MeV. The dotted line
(line 3) and the dash-dotted line (line 4) correspond to
$\bar\Lambda=750$ MeV and $\bar\Lambda=800$
MeV\footnote{$\bar\Lambda$ is the difference between the heavy
baryon mass and the heavy quark mass. This figure is in somewhat
analogue to that we achieved for the transition of a spin-1/2 baryon
to another spin-1/2 baryon. } respectively.

The dependence of the Isgur-Wise function $\xi_2$ on $\omega$  is
depicted in Fig.\ref{f3}(b). The dotted line (line 1) and the solid
line (line 2) are the results obtained in Scheme I and Scheme II
respectively. We observe that $\xi_2|_{\omega=1}=1/2$ in Scheme I is
consistent with that obtained in Ref.
\cite{Chow:1994ni,Cheng:1996cs}.

\begin{figure}[hhh]
\begin{center}
\scalebox{0.8}{\includegraphics{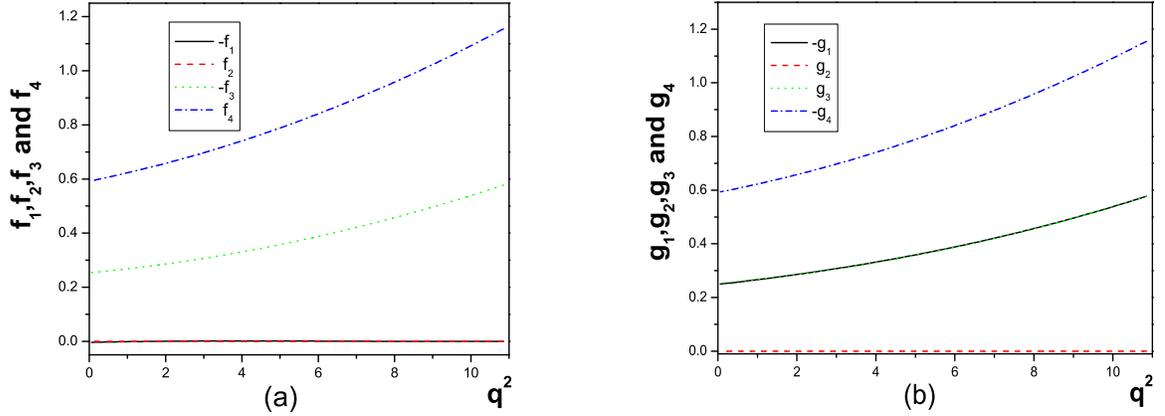}}
\end{center}
\caption{(a)  the form factors  $f_i\; (i=1,2,3,4)$and (b) the
form factors $g_i\; (i=1,2,3,4)$   in heavy quark limit in Scheme
I}\label{f41}
\end{figure}

Using the aforementioned relations between the form factors
$f_{i}\;(i=1,2,3,4)$ and $g_{i}\;(i=1,2,3,4))$ and the Isgur-Wise
functions $(\xi_1$ and $\xi_2)$, we obtain the form factors
$f_{i}\;(i=1,2,3,4)$ and $g_{i}\;(i=1,2,3,4)$ in the heavy quark
limit. According to the relations in Eq. (\ref{IWF}) $f_2$ and $g_2$ are
equal to 0.  $f_3$ nearly overlaps with $f_2$ and $g_3$ overlaps
with $g_4$.

By Fig. \ref{f21} and Fig. \ref{f41} we see that without taking
the heavy quark limit the absolute values of the the form factors
$f_{i}(i=1,2,3,4)$ and $g_1$  are slightly larger than those got
with the heavy quark limit. Especially the values of $f_2(q^2)$
with the heavy quark limit are exactly equal to 0 but the values
without the heavy quark limit slightly deviate from zero. One also
can find that the values of $g_2(q^2)$ without the heavy quark
limit are close to those with the heavy quark limit and the
absolute values of the the form factors $g_{3}$ and $g_{4}$
without the heavy quark limit are slightly larger than those with
the heavy quark limit at the small value of $q^2$. For the larger
$q^2$, $g_3(q^2)$ and $g_4(q^2)$ deviate from the values under the
heavy quark limit to a certain extent.

\subsection{$\Sigma_b\to \Sigma^*_c$ form factors and the Isgur-Wise
functions in Scheme II}

\begin{figure}
\begin{center}
\scalebox{0.8}{\includegraphics{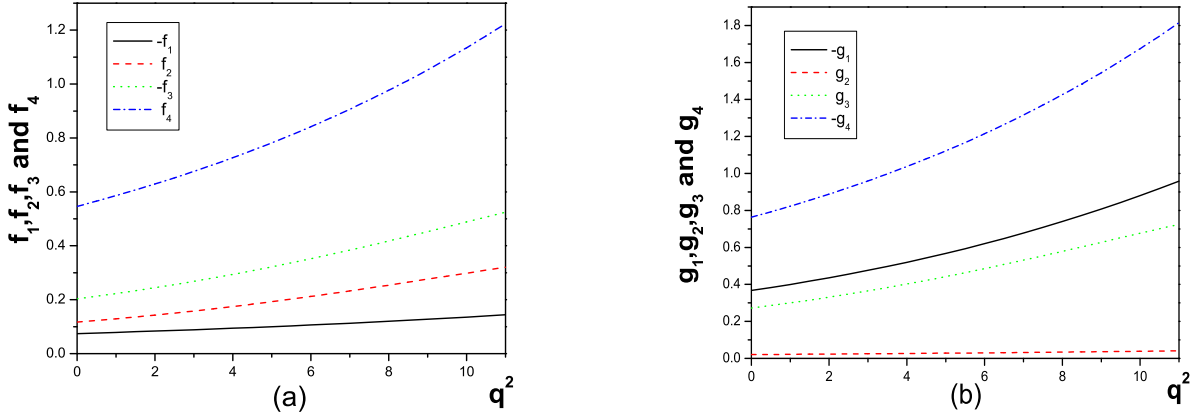}}
\end{center}
\caption{(a)  the form factors  $ f_{i}\;(i=1,2,3,4)$ and (b) the
form factors $g_{i}\;(i=1,2,3,4)$ in Scheme II }\label{f2}
\end{figure}

Now we repeat the calculation done in last subsection within the
framework of Scheme II. The fitted values of $a,~b$ and $F(0)$ in
the form factors $f_{i}\;(i=1,2,3,4)$ and $g_{i}\;(i=1,2,3,4)$ are
presented in Table \ref{Tab:t2}. The dependence of the form
factors on $q^2$ is depicted in Fig. \ref{f2}. Comparing Fig.
\ref{f2} and Fig. \ref{f21} one can find that $f_{i}\;(i=1,2,3,4)$
and $g_{i}\;(i=1,2,3,4)$ change more smoothly in Fig. \ref{f2},
the values of $f_{i}\;(i=1,2,3,4)$ in the two schemes are close to
each other and the values of $g_2$ in the two schemes are close to
0 but there are some apparent differences between the values of
$g_1, g_3$ and $g_4$ in the two schemes.
\begin{table}
\caption{The $\Sigma_b\to \Sigma^*_c$ form factors given in the
  three-parameter form in Scheme II.}\label{Tab:t2}
\begin{ruledtabular}
\begin{tabular}{cccc}
  $F$    &  $F(0)$ &  $a$  &  $b$ \\\hline
  $f_1$  &   -0.0744     &   0.967    & 0.755   \\
$f_2$  &   0.117     &   2.39    & 3.02   \\
  $f_3$  &    -0.203   &   2.13    &  2.52  \\
  $f_4$  &     0.546   &   1.37    &  1.02  \\
  $g_1$  &      -0.367    &     1.88 &  1.70  \\
  $g_2$  &      0.0205    &     1.19  &  1.29  \\
    $g_3$  &      0.272   &     2.31  & 2.91\\
  $g_4$  &      -0.763   &     1.54  & 1.18
\end{tabular}
\end{ruledtabular}
\end{table}

 In the
heavy quark limit, the Isgur-Wise function is fitted as
 \beq
\xi_1=1-2.08(\omega-1)+1.84(\omega-1)^2\\
\xi_2=0.409[1-2.75(\omega-1)+2.98(\omega-1)^2].
 \eeq

The dependence of the Isgur-Wise functions on $\omega$  is shown in
Fig.\ref{f3}.  $\xi_1$ in the two schemes and those given in Ref
.\cite{Ivanov:1996fj} are close to each other.

However from Fig.\ref{f3}(b), we observe that in Scheme II
$\xi_2|_{\omega=1}=0.41$ which is slightly lower than $1/2$
\cite{Chow:1994ni,Cheng:1996cs}. In fact under the heavy quark
limit the mass of the heavy quark does not exist in the
wavefunction, but the mass of the light diquark remains(Eq.
(\ref{a1a2II})). Therefore the theoretical evaluation on the
transition rate may weakly depend on its mass. For calculating
hadronic matrix elements in terms of a concrete model with one or
several parameters which are not determined by any underlying
theory yet, we let $m_{{[ud]}_V}$ and $\beta^\infty$ vary within
reasonable ranges and see how much the numerical results depend on
it(them). In fact, it is the strategy which are generally adopted
for model-dependent phenomenological studies.

The resultant figures show that $\xi_1$ almost does not change with
respect to those variations, but the intercept $\xi_2|$ at
${\omega=1}$ is not the same  for different values of $m_{{[ud]}_V}$
and $\beta^\infty$. As a matter of fact the situation is exactly the
same as for transitions of a spin-1/2 baryon into another 1/2-baryon
under the heavy quark limit. Thus for simplicity let us recall
what we gave in our earlier work as: $\xi_1|_{\omega=1}=1$
and $\xi_2|_{\omega=1}=0.47$ as $m_{{[ud]}_V}=0.5$ GeV and
$\beta^\infty=0.4$. It is easy to understand that even though
non-zero $m_{[ud]_{_V}}$ breaks the heavy quark symmetry
$SU_f(2)\otimes SU_s(2)$, the violation is still rather small.
Therefore,  one still can use the simplified expressions with only
two Isgur-Wise functions to approach the hadronic matrix elements
for either a spin-1/2 to spin-1/2 transition or a spin-1/2 to
spin-3/2 transition under the limit.

\begin{figure}[hhh]
\begin{center}
\scalebox{0.8}{\includegraphics{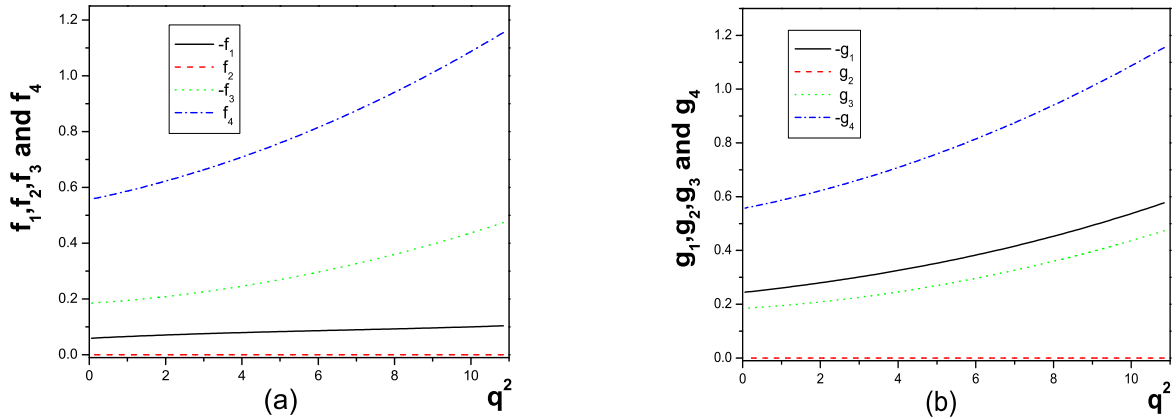}}
\end{center}
\caption{ form factors in heavy quark limit in Scheme
II}\label{f4}
\end{figure}

Using the aforementioned relations between the form factors
$f_{i}(i=1,2,3,4)$ and $g_{i}(i=1,2,3,4))$ and the Isgur-Wise
functions $(\xi_1$ and $\xi_2)$, we obtain the form factors
$f_{i}(i=1,2,3,4)$ and $g_{i}(i=1,2,3,4)$ in the heavy quark
limit. From the Fig. \ref{f2} and Fig. \ref{f4} we observe that
the absolute values of the the form factors $f_{i}(i=1,2,3,4)$ and
$g_{i}(i=1,2,3,4)$ without taking the heavy quark limit are
slightly larger than those with the heavy quark limit at the same
$q^2$. Especially the value of $f_2(q^2)$ with the heavy
quark limit is exactly equal to 0 but the value without the heavy
quark limit slightly deviates from zero.

\subsection{Semi-leptonic decay of $\Sigma_b \to
\Sigma_c^* +l\bar{\nu}_l$}

\begin{figure}[hhh]
\begin{center}
\scalebox{0.8}{\includegraphics{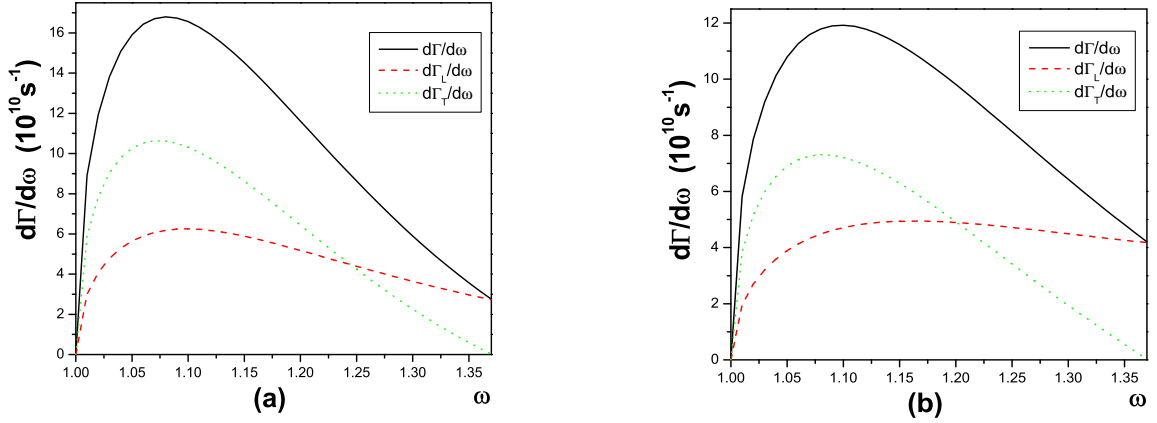}}
\end{center}
\caption{ Differential decay rates $d\Gamma/d\omega$ for the decay
$\Sigma_b \to \Sigma_c^* l\bar{\nu}_l$ in Scheme I(a) with heavy
quark limit; (b) without heavy quark limit}\label{f51}
\end{figure}

\begin{figure}[hhh]
\begin{center}
\scalebox{0.8}{\includegraphics{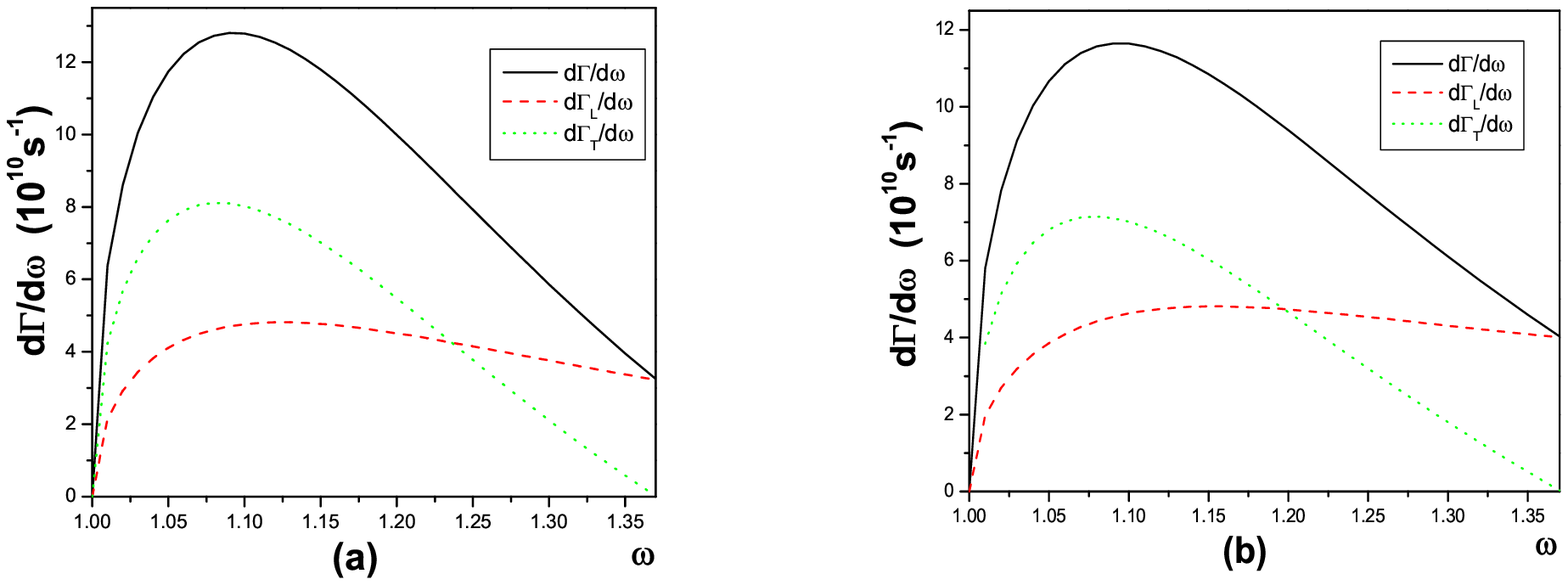}}
\end{center}
\caption{ Differential decay rates $d\Gamma/d\omega$ for the decay
$\Sigma_b \to \Sigma_c^* l\bar{\nu}_l$ in Scheme II(a) with heavy
quark limit; (b) without heavy quark limit}\label{f52}
\end{figure}

Using the form factors obtained in last subsection, we evaluate
the rate of $\Sigma_b \to \Sigma_c^* l\bar{\nu}_l$ in two cases:
with and without taking the heavy quark limit.  We list our
predictions in table \ref{Tab:t4}. The numerical results depend on
the light diquark mass, even though not very sensitively. In Table
 \ref{Tab:t4}, as discussed above, with the same strategy, we let the diquark
mass and parameters $\beta_{b[ud]}$, $\beta_{c[ud]}$ fluctuate up
to 10\%, and the corresponding changes are listed.

It is interesting to study a ratio of the longitudinal
differential rate to the transverse one  (which are integrated
over $\omega$ and the ration $R$ is defined in the appendix),
since it may provide more information about the model. $R\sim 1$
would imply the spatial distribution to be approximately
uniform. Because $R$ is more
sensitive to details of the employed models, a comparison of the
theoretical prediction with the data which will be available at
LHCb, can help to further constrain the model
parameters.

We list our numerical results of $R$ in In Tab.\ref{Tab:t4},
meanwhile the predictions achieved with other approaches
\cite{Ebert:2006rp} are also presented. One notices from
Tab.\ref{Tab:t4} that our results in the two schemes  are close to
that predicted by Ref.\cite{Ebert:2006rp}. In Scheme I the total
width without heavy quark limit is larger than that with   heavy
quark limit apparently. In Scheme II the total width without heavy
quark limit is close to that with heavy quark limit.

\begin{table}
\caption{The width (in unit $10^{10}{\rm s}^{-1}$)   of
$\Sigma_b\to \Sigma^*_c l\bar{\nu}_l$ with
$m_{[ud]_{_V}}=770\pm77$MeV, $\beta_{c[ud]}=0.45\pm0.05$ and
$\beta_{b[ud]}=0.50\pm0.05 $.}\label{Tab:t4}
\begin{ruledtabular}
\begin{tabular}{c|ccccc}
    &  $\Gamma$ &  $\Gamma_L$  &  $\Gamma_T$ & $R$    \\\hline
 this work in Scheme I\footnote{ with the heavy quark limit}
   &    3.27$\pm0.34$    & 1.62$\pm0.20$ & 1.65$\pm0.15$& 0.984$\pm0.033$  \\
  this work in Scheme I\footnote{without the heavy quark limit}
    &  4.03$\pm0.40$  & 1.76$\pm0.16$  & 2.27$\pm0.24$& 0.775$\pm0.013$  \\
 this work in Scheme II$^a$
   &    3.17$\pm$0.30    & 1.58$\pm$0.16 & 1.59$\pm$0.13& 0.994$\pm$0.024  \\
  this work in Scheme II$^b$
    &  3.34$\pm$0.26   & 1.51$\pm$0.13  & 1.83$\pm$0.13& 0.825$\pm0.024$  \\
  relativistic quark model$^a$\cite{Ebert:2006rp} &   3.23      & 1.61  & 1.62  & 0.99\\
  relativistic three-quark model$^a$\cite{Ivanov:1996fj}& 4.56   &  2.49 & 2.07 & 1.20 \\
the Bethe-Salpeter approach$^a$\cite{Ivanov:1998ya}&3.75 & - &-&-
\end{tabular}
\end{ruledtabular}
\end{table}

The dependence of the differential decay rate of $\Sigma_b \to
\Sigma_c^* l\bar{\nu}_l$ on $\omega$ is depicted in Fig. \ref{f51}
and Fig. \ref{f52} for Scheme I and II respectively. Fig.
\ref{f51} (Fig. \ref{f52} ) (a) and (b) are corresponding to the
results with or without taking the heavy quark limits, and the two
figures in Fig. \ref{f52} (Scheme II) are very similar to Fig.
\ref{f51} (Scheme I) except the position of the crossing points of
the dashed and the dotted lines. In addition the differential decay
rate of $\Sigma_b \to \Sigma_c^* l\bar{\nu}_l$ on $\omega$  in the
two schemes with heavy quark limit are close to each other but
there exists a difference between the results  in the two schemes
when heavy quark limit is not taken into account.

\subsection{Non-leptonic decays of ${\Sigma_b}\to\Sigma_c^*+ X$}

From the theoretical aspects, calculating the concerned quantities
of the non-leptonic decays seems to be more complicated than the
semi-leptonic ones, but can still shed lights on the properties of
the chosen model. Our theoretical framework is based on the
factorization assumption, namely the hadronic transition matrix
element is factorized into a product of two independent matrix
elements of currents,
\begin{eqnarray}\label{s0}
&& \la \Sigma_{c}^*(P',S_z')X \mid \mathcal{H} \mid \Sigma_{b}(P,S_z) \ra  \nonumber \\
 &=&\frac{G_FV_{bc}V^*_{qq'}}{\sqrt{2}}\la X \mid
\bar{q'} \gamma^{\mu} (1-\gamma_{5}) q \mid 0\ra\la
\Sigma_{c}^*(P',S_z') \mid \bar{c} \gamma^{\mu} (1-\gamma_{5}) b
\mid \Sigma_{b}(P,S_z) \ra,
\end{eqnarray}
where the part $\la X \mid \bar{q'} \gamma^{\mu} (1-\gamma_{5}) q
\mid 0\ra$ is determined by the decay constant of meson $X$ and
the transition matrix element $\la \Sigma_{c}^*(P',S_z') \mid
\bar{c} \gamma^{\mu} (1-\gamma_{5}) b \mid \Sigma_{b}(P,S_z) \ra$
was studied in the previous sections.
Since the decay $\Sigma_b^0\to\Sigma_c^*+X^-$ is the so-called
color-favored transition, the factorization should be a good
approximation. The study on these non-leptonic decays can check
the validity degree of the obtained form factors for the heavy
bottomed baryon system.

Our numerical results are shown in Tab.\ref{Tab:t6}, where the
uncertainties originate from the variation of $m_{[ud]_{_V}}$ and
$\beta$ which are allowed to fluctuate by 10$\%$. The effective
Wilson coefficient $a_1$ is set as 1 and the meson decay constants
take the same values given in Ref.\cite{Ke:2007tg}.

Some comments are made:

(1) The ratio $\frac{BR(\Sigma_b\to\Sigma_c^*
l^-\bar\nu_l)}{BR(\Sigma_b \to\Sigma_c^*\pi^-)}$ without the heavy
quark limit is  $34.44\pm6.56$ (Scheme I) or $23.69\pm5.37$
(Scheme II) which will be experimentally tested and the
consistency would tell us which scheme is the more realistic one.

(2)The numerical results given in Table IV indicate that the decay
rate $\Gamma(\Sigma_b\to\Sigma_c^* V)$
is 2 to 3 times larger than
$\Gamma(\Sigma_b\to\Sigma_c^* P)$ where P and V are pseudoscalar
and vector mesons with the same quark flavors.

\begin{table}
\caption{widths (in unit $10^{10}{\rm s}^{-1}$) of non-leptonic
decays $\Sigma_b\to\Sigma_c^* X$ with the light diquark mass
$m_{[ud]_{_V}}=770\pm77$ MeV and
$\beta=0.50\pm0.05$.}\label{Tab:t6}
\begin{ruledtabular}
\begin{tabular}{ccccc}
 &Scheme I\footnote{ without the heavy quark limit} & Scheme II$^a$& Scheme I\footnote{ with the heavy quark limit}
  &Scheme II$^b$ \\\hline
  $\Sigma_b^0\to\Sigma_c^* \pi^-$ & 0.117$\pm0.019$  & $0.141\pm$0.030 &$0.138\pm0.030$&
  $0.132\pm0.025$
                             \\\hline
  $\Sigma_b^0\to\Sigma_c^* \rho^-$ &$0.391\pm0.060$&
  $0.447\pm0.086$&0.430$\pm0.091$
                        & $0.411\pm0.082$   \\\hline
  $\Sigma_b^0\to\Sigma_c^* K^-$   &0.0094$\pm0.0015$ &
  $0.0113\pm0.0023$&$0.0109\pm0.0024$
                        & $0.0104\pm0.0019$       \\\hline
  $\Sigma_b^0\to\Sigma_c^* K^{*-}$ & $0.0211\pm0.0032$&
  $0.0237\pm0.0045$&$0.0227\pm0.0047$
                        & $0.0217\pm0.0039$   \\\hline
  $\Sigma_b^0\to\Sigma_c^* a_1^-$ & 0.457$\pm0.064$ &
  $0.486\pm0.083$&0.459$\pm0.141$
                        & $0.437\pm0.076$   \\\hline
  $\Sigma_b^0\to\Sigma_c^* D^-$  &0.0163$\pm0.0016$ & $0.0169\pm0.0025$ & $0.0144\pm0.0023$ &
  $0.0139\pm0.0020$
                       \\\hline
  $\Sigma_b^0\to\Sigma_c^* D^{*-}$&0.0535$\pm0.0057$ &
  $0.0495\pm0.0054$&$0.0452\pm0.0067$
                        & $0.0431\pm0.0058$   \\\hline
  $\Sigma_b^0\to\Sigma_c^* {D_s^-}$ &0.423$\pm0.038$ &
  $0.433\pm0.060$&0.364$\pm0.260$
                        & $0.351\pm0.0048$   \\\hline
  $\Sigma_b^0\to\Sigma_c^* D^{*-}_s$&1.25$\pm0.13$ &$1.13\pm0.12$&1.03$\pm0.14$
                        & $0.99\pm0.13$
\end{tabular}
\end{ruledtabular}
\end{table}

\subsection{Comparison of the results on $\Sigma_b\to\Sigma_c$+other(s) with  those on  $\Sigma_b\to\Sigma_c^*$+other(s)}
\begin{table}
\caption{widths (in unit $10^{10}{\rm s}^{-1}$)  of some decays
$\Sigma_b\to\Sigma_c(\Sigma_c^*)$+other(s)  without the heavy
quark limit in Scheme II.}\label{Tab:t7}
\begin{ruledtabular}
\begin{tabular}{ccc}
  & $B$ represents $\Sigma_c$
  & $B$ represents $\Sigma_c^*$\\\hline
  $\Sigma_b \to B l\bar{\nu}_l$&   1.60$\pm$0.28             &  3.34$\pm$0.26        \\\hline
  $\Sigma_b^0\to B  \pi^-$  & $0.140\pm0.037$
                        & $0.141\pm$0.030     \\\hline
  $\Sigma_b^0\to B \rho^-$ & $0.396\pm0.091$
                        & $0.447\pm0.086$    \\\hline
  $\Sigma_b^0\to B K^-$    & $0.0115\pm0.0030$
                        &$0.0113\pm0.0023$     \\\hline
  $\Sigma_b^0\to B K^{*-}$  & $0.0204\pm0.0041$
                        & $0.0237\pm0.0045$     \\\hline
  $\Sigma_b^0\to B a_1^-$  & $0.369\pm0.068$
                        & $0.486\pm0.083$  \\\hline
  $\Sigma_b^0\to B D_s^-$  & $0.727\pm0.150$
                        &  $0.433\pm0.060$    \\\hline
  $\Sigma_b^0\to B D^{*-}_s$& $0.558\pm0.067$
                        & $1.13\pm0.12$   \\\hline
  $\Sigma_b^0\to B D^-$  & $0.0266\pm0.0056$
                        &$0.0169\pm0.0025$    \\\hline
  $\Sigma_b^0\to B {D^*}^-$& $0.0261\pm0.0035$
                        &$0.0495\pm0.0054$
\end{tabular}
\end{ruledtabular}
\end{table}
Since the only difference between $\Sigma_c$ and $\Sigma_c^*$ is
their total spins, it is natural to expect that there possibly
exists a simple relation between the decay rates of
$\Sigma_b\to\Sigma_c+X$ and $\Sigma_b\to\Sigma_c^*+X$.
Now let us study the  relation
in terms of our numerical results.

From table \ref{Tab:t7} one can notice:

(1) The ratio $\frac{BR(\Sigma_b\to\Sigma_c^*
l^-\bar\nu_l)}{BR(\Sigma_b \to\Sigma_cl^-\bar\nu_l)}$  is about 2.

(2)The theoretically predicted widths of $\Sigma_b\to\Sigma_c$+light scalar and
$\Sigma_b\to\Sigma_c^*$+light scalar  are
nearly equal.

(3)The ratio of B($\Sigma_b\to\Sigma_c^*$+light vector) over
B($\Sigma_b\to\Sigma_c$+light vector) sways from 1 to 2, which is
roughly consistent with the value 1.5 required by the $SU_s(2)$
symmetry in HQET.

It is noted that a factor 2
was missing in the formula for the transition
$\frac{1}{2}\to\frac{1}{2}+V$ given by\cite{Cheng:1996cs}
\footnote{We have discussed this issue with the authors of
Ref.\cite{Cheng:1996cs}, and then by having carefully checked  this formula
they agreed with us.
 }. We used it to calculate the rate of
$\Sigma_b\to\Sigma_c+V$ in Ref.\cite{Ke:2012wa}. In this work, by
noticing that mistake, we rewrite the formula by adding up the
missed factor and carry out the corresponding calculations. In
table \ref{Tab:t7}, the theoretical predictions of
$\Sigma_b\to\Sigma_c+V$ are made in terms of the corrected
formula.



\section{Conclusions and discussions}
In this paper, we explore the $\Sigma_b\to\Sigma^*_c$ transitions in
the light front quark model. We calculate the widths of the
semi-leptonic decays and non-leptonic two-body decays of
$\Sigma_b\to\Sigma^*_c $ where the quark-diquark picture is employed
so that the three-body inner structure of the baryons is reduced  into a two-body one.

Since there exists an axial diquark in $\Sigma_b$ and $\Sigma^*_c$,
one should expect its polarization vector in the vertex function
is somehow momentum-dependent. The polarization vector deduced
from the CG coefficients uniquely depends on the momentum of the
baryon. With this momentum dependence, we name it as the Scheme I
for later calculation. Alternatively, based on a physical
consideration, we would set the Scheme II  where the polarization
vector of the diquark depends on its own momentum which may
respect the identity $p_2\cdot\epsilon\equiv 0$,
since in our model, the diquark is
on its mass shell. Then, we calculate the eight form factors and
two generalized Isgur-Wise functions $\xi_1$ and $\xi_2$ under the
heavy quark limit in the two schemes. The form factors
$f_i\;(i=1,2,3,4)$ and $g_i\;(i=1,2,3,4)$ in Scheme II change more
smoothly than those in Scheme I. The values of $f_i\;(i=1,2,3,4)$
, $g_1$ and $g_2$ in the two schemes are close to each other but
there are some apparent differences for $g_3$ and $g_4$ in the two
schemes. $\xi_1$'s in the two schemes are close to each other and
consistent with the results in references. However there is a
discrepancy between $\xi_2$'s in the two schemes.
$\xi_2|_{\omega=1}$ in Scheme I is exactly 1/2 but
$\xi_2|_{\omega=1}$ in Scheme II is slightly lower than 1/2
predicted by the large $N_c$ theory. The deviation may be due to
the non-zero mass of the light constituents in hadrons ( baryon).
We also evaluate the form factors $f_i\;(i=1,2,3,4)$ and
$g_i\;(i=1,2,3,4)$ in the heavy quark limit in terms of the
Isgur-Wise functions $\xi_1$ and $\xi_2$ in the two schemes and
there are not apparently differences between them. Applying the
form factors derived in the framework of the LFQM we evaluate the
semi-leptonic decay rates of $\Sigma_b\to\Sigma^*_c$ with and
without taking the heavy quark limit. The results in both cases do
not decline much from each other, moreover, our numerical results
are qualitatively consistent with that estimated in terms of
different approaches.

It is noted that the Scheme I retains
apparent Lorentz invariance for the vertex function
whereas, even though the Scheme II seems to be more physical, the
cost we pay is that the explicit Lorentz invariance is lost. This is
a more profound question which we are going to address in our
following work.

Since the RUN-II of the LHC is operating well and a remarkable
number of data on $\Sigma_b$( production and decay) is being
accumulated by the LHCb collaboration, we have all confidence that
in near future the rates and even the details of  various decay
modes would be accurately measured, so theorists will have a great
opportunity to testify all available models and re-fix relevant
model parameters.

\section*{Acknowledgement}

This work is supported by the National Natural Science Foundation
of China (NNSFC) under the contract No. 11375128 and 11675082. We
thank Prof. Hai-Yang Cheng, Dr. Xian-Wei Kang and Dr. Fanrong Xu
for helpful discussions on the missing factor. We also benefit
from the pleasant communication with Prof. Chun-Khiang Chua, Dr.
Wei Wang and Dr. Yan-Liang Shi on the relevant issues.

\appendix

\section{Semi-leptonic decays of
 $\Sigma_Q \to \Sigma^*_{Q'} l\bar\nu_l$ }

The helicity amplitudes are expressed in terms of the form factors
for $\Sigma_Q\to\Sigma^*_{Q'}$ \cite{Korner:1991ph,Bialas:1992ny}
\begin{eqnarray}
  \label{eq:haad}
  H^{V,A}_{1/2,\, 0}&=&\mp\frac1{\sqrt{q^2}}\frac{2}{\sqrt3}
\sqrt{M_{\Sigma_Q}M_{\Sigma^*_{Q'}}(w\mp
  1)}[(M_{\Sigma_Q}w-M_{\Sigma^*_{Q'}}){\cal N}^{V,A}_4(w)\cr
&&\mp (M_{\Sigma_Q}\mp M_{\Sigma^*_{Q'}})(w\pm1){\cal
N}^{V,A}_1(w) +M_{\Sigma^*_{Q'}}(w^2-1){\cal N}^{V,A}_2(w)\cr
&&+M_{\Sigma_Q}(w^2-1){\cal N}^{V,A}_3(w)],\cr
 H^{V,A}_{1/2,\, 1}&=&\sqrt{\frac23}\sqrt{M_{\Sigma_Q}M_{\Sigma^*_{Q'}}(w\mp
  1)}[{\cal N}^{V,A}_4(w)-2(w\pm 1){\cal N}^{V,A}_1(w)],\cr
H^{V,A}_{3/2,\, 1}&=&\mp\sqrt{2M_{\Sigma_Q}M_{\Sigma^*_{Q'}}(w\mp
  1)}{\cal N}^{V,A}_4(w),
\end{eqnarray}
where again the upper(lower)  sign corresponds  to $V(A)$ and
${\cal
  N}^V_i\equiv g_i$, ${\cal N}^A_i\equiv f_i$ ($i=1,2,3,4$).
The remaining helicity amplitudes  can be obtained using the
relation
$$H^{V,A}_{-\lambda',\,-\lambda_W}=\mp H^{V,A}_{\lambda',\, \lambda_W}.$$
Partial differential decay rates can be represented in the
following form
\begin{eqnarray}
  \label{eq:darlt}
 \frac{d\Gamma_T}{dw}&=&\frac{G_F^2}{(2\pi)^3} |V_{QQ'}|^2\frac{q^2
   M_{\Sigma^*_{Q'}}^2\sqrt{w^2-1}}{12M_{\Sigma_Q}} [|H_{1/2,\, 1}|^2+
   |H_{-1/2,\, -1}|^2+|H_{3/2,\, 1}|^2+
   |H_{-3/2,\, -1}|^2],\cr
\frac{d\Gamma_L}{dw}&=&\frac{G_F^2}{(2\pi)^3} |V_{QQ'}|^2\frac{q^2
   M_{\Sigma^*_{Q'}}^2\sqrt{w^2-1}}{12M_{\Sigma_Q}} [|H_{1/2,\, 0}|^2+
   |H_{-1/2,\, 0}|^2].
\end{eqnarray}

The ratio of the longitudinal to transverse decay rates $R$ is
defined by
 \beq
 R=\frac{\Gamma_L}{\Gamma_T}=\frac{\int_1^{\Sigma_{\rm
     max}}d\Sigma~q^2~p_c\left[ |H_{\frac{1}{2},0}|^2+|H_{-\frac{1}{2},0}|^2
     \right]}{\int_1^{\Sigma_{\rm max}}d\Sigma~q^2~p_c
     \left[ |H_{\frac{1}{2},1}|^2+|H_{-\frac{1}{2},-1}|^2\right]}.
 \eeq

\section{the form factor of
 $\Sigma_Q \to \Sigma_{Q^*}$ }

 $  \bar u(\bar P,S_z)\gamma^{\mu}\bar P^\beta
{u}_{\beta}(\bar P',S'_z)$, $\bar u(\bar P,S_z)\bar P'^{\mu}\bar
P^\beta{u}_{\beta}(\bar P',S'_z)$ , $\bar u(\bar P,S_z) \bar
P^{\mu}\bar P^\beta{u}_{\beta}(\bar P',S'_z)$, $\bar u(\bar P,S_z)
g^{\mu \beta}{u}_{\beta}(\bar P',S'_z)$ are multiplied to the
right side of Eq.(\ref{s00}), then
\begin{eqnarray}\label{s01}
 F_1&=&\int\{d^3 \tilde p_2\}\frac{\phi_{\Sigma_{c}^*}^*(x',k'_{\perp})
  \phi_{\Sigma_b}(x,k_{\perp})}{2\sqrt{3p^+_1p'^+_1(p_1\cdot \bar{P}+m_1M_0)
 (p'_1\cdot \bar{P'}+m'_1M'_0)}}{\rm Tr}\big\{{u}_\beta(\bar{P'},S'_z)\bar{u}_\alpha(\bar{P'},S'_z)\nonumber \\
  &&\times  [\varepsilon^{\alpha}(\bar{P'},\sigma)]
  (p_1\!\!\!\!\!\slash'+m'_1)\gamma^{\mu}(-\gamma_{5})
  (p_1\!\!\!\!\!\slash+m_1) [-\gamma_5\varepsilon^*\!\!\!\!\!\slash(\bar P,\sigma)] u(\bar{P},S_z)\bar u(\bar{P},S_z)\gamma_{\mu}\bar
  P^\beta\big\},\\
   F_2&=&\int\{d^3 \tilde p_2\}\frac{\phi_{\Sigma_{c}^*}^*(x',k'_{\perp})
  \phi_{\Sigma_b}(x,k_{\perp})}{2\sqrt{3p^+_1p'^+_1(p_1\cdot \bar{P}+m_1M_0)
 (p'_1\cdot \bar{P'}+m'_1M'_0)}}{\rm Tr}\big\{{u}_\beta(\bar{P'},S'_z)\bar{u}_\alpha(\bar{P'},S'_z)\nonumber \\
  &&\times  [\varepsilon^{\alpha}(\bar{P'},\sigma)]
  (p_1\!\!\!\!\!\slash'+m'_1)\gamma^{\mu}(-\gamma_{5})
  (p_1\!\!\!\!\!\slash+m_1) [-\gamma_5\varepsilon^*\!\!\!\!\!\slash(\bar P,\sigma)] u(\bar{P},S_z)\bar u(\bar{P},S_z)\bar P'_{\mu}\bar
  P^\beta\big\},\\ F_3&=&\int\{d^3 \tilde p_2\}\frac{\phi_{\Sigma_{c}^*}^*(x',k'_{\perp})
  \phi_{\Sigma_b}(x,k_{\perp})}{2\sqrt{3p^+_1p'^+_1(p_1\cdot \bar{P}+m_1M_0)
 (p'_1\cdot \bar{P'}+m'_1M'_0)}}{\rm Tr}\big\{{u}_\beta(\bar{P'},S'_z)\bar{u}_\alpha(\bar{P'},S'_z)\nonumber \\
  &&\times  [\varepsilon^{\alpha}(\bar{P'},\sigma)]
  (p_1\!\!\!\!\!\slash'+m'_1)\gamma^{\mu}(-\gamma_{5})
  (p_1\!\!\!\!\!\slash+m_1) [-\gamma_5\varepsilon^*\!\!\!\!\!\slash(\bar P,\sigma)] u(\bar{P},S_z)\bar u(\bar{P},S_z)\bar P_{\mu}\bar
  P^\beta\big\},\\
   F_4&=&\int\{d^3 \tilde p_2\}\frac{\phi_{\Sigma_{c}^*}^*(x',k'_{\perp})
  \phi_{\Sigma_b}(x,k_{\perp})}{2\sqrt{3p^+_1p'^+_1(p_1\cdot \bar{P}+m_1M_0)
 (p'_1\cdot \bar{P'}+m'_1M'_0)}}{\rm Tr}\{{u}_\beta(\bar{P'},S'_z)\bar{u}_\alpha(\bar{P'},S'_z)\nonumber \\
  &&\times  [\varepsilon^{\alpha}(\bar{P'},\sigma)]
  (p_1\!\!\!\!\!\slash'+m'_1)\gamma^{\mu}(-\gamma_{5})
  (p_1\!\!\!\!\!\slash+m_1) [-\gamma_5\varepsilon^*\!\!\!\!\!\slash(\bar P,\sigma)] u(\bar{P},S_z)\bar u(\bar{P},S_z)g_{\mu}^\beta \},
\end{eqnarray}

 $  \bar u(\bar P,S_z)\gamma^{\mu}\bar P^\beta
{u}_{\beta}(\bar P',S'_z)$, $\bar u(\bar P,S_z)\bar P'^{\mu}\bar
P^\beta{u}_{\beta}(\bar P',S'_z)$ , $\bar u(\bar P,S_z) \bar
P^{\mu}\bar P^\beta$ ${u}_{\beta}(\bar P',S'_z)$, $\bar u(\bar
P,S_z) g^{\mu \beta}{u}_{\beta}(\bar P',S'_z)$ are timed on the
right side of Eq.(\ref{s2}),
\begin{eqnarray}\label{s21}
F_1 &=&{\rm Tr}\{{u}_\beta(\bar{P'},S'_z) \bar{u}_{\alpha}(\bar
P',S'_z) \left[ \gamma^{\mu}\bar P^\alpha
\frac{f_{1}(q^{2})}{M_{\Sigma_b}}
 + \frac{f_{2}(q^{2})}{M_{\Sigma_b}M_{\Sigma_b}}\bar P^\alpha \bar P^\mu+\frac{f_{3}(q^{2})}{M_{\Sigma_b}M_{\Sigma^*_c}}\bar P^\alpha \bar P'^\mu
 +f_4g^{\alpha \mu}
 \right] \nonumber\\&&u(\bar P,S_z)\bar u(\bar{P},S_z)\gamma_{\mu}\bar
  P^\beta \},\\F_2 &=&{\rm Tr}\{{u}_\beta(\bar{P'},S'_z) \bar{u}_{\alpha}(\bar
P',S'_z) \left[\gamma^{\mu}\bar P^\alpha
\frac{f_{1}(q^{2})}{M_{\Sigma_b}}
 + \frac{f_{2}(q^{2})}{M_{\Sigma_b}M_{\Sigma_b}}\bar P^\alpha \bar P^\mu+\frac{f_{3}(q^{2})}{M_{\Sigma_b}M_{\Sigma^*_c}}\bar P^\alpha \bar P'^\mu
 +f_4g^{\alpha \mu}
 \right] \nonumber\\&&u(\bar P,S_z)\bar u(\bar{P},S_z)\bar P'_{\mu}\bar
  P^\beta \},\\F_2 &=&{\rm Tr}\{{u}_\beta(\bar{P'},S'_z) \bar{u}_{\alpha}(\bar
P',S'_z) \left[\gamma^{\mu}\bar P^\alpha
\frac{f_{1}(q^{2})}{M_{\Sigma_b}}
 + \frac{f_{2}(q^{2})}{M_{\Sigma_b}M_{\Sigma_b}}\bar P^\alpha \bar P^\mu+\frac{f_{3}(q^{2})}{M_{\Sigma_b}M_{\Sigma^*_c}}\bar P^\alpha \bar P'^\mu
 +f_4g^{\alpha \mu}
 \right] \nonumber\\&&u(\bar P,S_z)\bar u(\bar{P},S_z)\bar P_{\mu}\bar
  P^\beta \},
  \\F_4 &=&{\rm Tr}\{{u}_\beta(\bar{P'},S'_z) \bar{u}_{\alpha}(\bar
P',S'_z) \left[ \gamma^{\mu}\bar P^\alpha
\frac{f_{1}(q^{2})}{M_{\Sigma_b}}
 + \frac{f_{2}(q^{2})}{M_{\Sigma_b}M_{\Sigma_b}}\bar P^\alpha \bar P^\mu+\frac{f_{3}(q^{2})}{M_{\Sigma_b}M_{\Sigma^*_c}}\bar P^\alpha \bar P'^\mu
 +f_4g^{\alpha \mu}
 \right] \nonumber\\&&u(\bar P,S_z)\bar u(\bar{P},S_z)g_{\mu}^\beta\}.
\end{eqnarray}
Then by solving these equations one can obtain the expressions of
$f_1$, $f_2$, $f_3$, $f_4$.

\end{document}